\documentclass[10pt,conference,letterpaper]{IEEEtran}
\usepackage{times,amsmath,epsfig}
\usepackage{threeparttablex,multirow,booktabs} 
\usepackage[noend]{algpseudocode}
\usepackage[ruled,vlined,boxed,linesnumbered]{algorithm2e}
\usepackage{subcaption}
\usepackage{amsfonts}
\usepackage{mathrsfs}
\usepackage[colorlinks=black,linkcolor=black,anchorcolor=black,citecolor=black]{hyperref}
\graphicspath{{figures/}{figure/}{pictures/}{picture/}{pic/}{pics/}{image/}{images/}{img/}}

\title{\textsc{DancingLines}: An Analytical Scheme to Depict Cross-Platform Event Popularity}
\author{%
	{Tianxiang Gao{\small $^{\#}$}, Weiming Bao{\small $^{\#}$}, Jinning Li{\small $^{\#}$}, Xiaofeng Gao{\small $^{\#}$}, Boyuan Kong{$^{*}$}, Yan Tang{$^{\dagger}$}, Guihai Chen{\small $^{\#}$}, Xuan Li{$^{\ddagger}$}}\\%
	\fontsize{10}{10}\selectfont\itshape
	$^{\#}$\,Department of Computer Science and Engineering, Shanghai Jiao Tong University, Shanghai, China\\
	\fontsize{9}{9}\selectfont\ttfamily\upshape
	\{gtx9726, wm\_bao, lijinning, gaoxiaofeng, chen-gh\}@sjtu.edu.cn\\
	\fontsize{10}{10}\selectfont\rmfamily\itshape
	$^{*}$\,Department of Electrical Engineering and Computer Science, University of California, Berkeley, CA, USA\\
	\fontsize{9}{9}\selectfont\ttfamily\upshape
	boyuan\_kong@berkeley.edu\\
	\fontsize{10}{10}\selectfont\rmfamily\itshape
	$^{\dagger}$\,College of Computer and Information, Hohai University, Nanjing, China\\
	\fontsize{9}{9}\selectfont\ttfamily\upshape
	tangyan@hhu.edu.cn\\
	
	\fontsize{10}{10}\selectfont\rmfamily\itshape
	$^{\ddagger}$\,Baidu, Inc., Beijing, China\\
	\fontsize{9}{9}\selectfont\ttfamily\upshape
	xli@baidu.com
}
\begin{document}
	\maketitle
	\begin{abstract}

		Nowadays, events usually burst and are propagated online through multiple modern media like social networks and search engines.
		There exists various research discussing the event dissemination trends on individual medium, while few studies focus on event popularity analysis from a cross-platform perspective.
		Challenges come from the vast diversity of events and media, limited access to aligned datasets across different media and
		the great deal of noise in the datasets.


		In this paper, we design \textsc{DancingLines}, an innovative scheme that captures and quantitatively analyzes event popularity between pairwise text media. It contains two models: TF-SW, a semantic-aware popularity quantification model, based on an integrated weight coefficient leveraging Word2Vec and TextRank;
		and $\omega$DTW-CD, a pairwise \textit{event popularity time series} alignment model matching
		different event phases adapted from Dynamic Time Warping.
		We also propose three metrics to interpret event popularity trends between pairwise social platforms.
		


		Experimental results on eighteen real-world event datasets from an influential social network and a popular search engine validate the effectiveness and applicability of our scheme. \textsc{DancingLines} is demonstrated to possess broad application potentials for discovering knowledge of various aspects related to events and different media.

	\end{abstract}

	%
	\section{Introduction}
	\label{sec:intro}
	With the development of Internet technology, in recent years, the primary media for information propagation have been shifting to online media, such as social networks, search engines, web portals, etc. A vast number of studies have been conducted to analyze the event disseminations comprehensively on a single medium~\cite{DBLP:conf/icde/LiLKC12,DBLP:conf/kdd/LinWHY13,DBLP:conf/icdm/WangTYLMCHH15,DBLP:journals/vldb/Zhou014}.
	In fact, an event is less likely to be captured only by single platform, and popular events are usually disseminated on multiple media. 
	Thus, depicting and analyzing event popularity across different platforms plays a vital role in tracking the public concerns and understanding the event disseminations. Moreover, natural features of events, properties and the interactions of media can be revealed by this analysis.

	However, few works have been done to depict characters of event popularity from a cross-platform perspective.
	Challenges exist in multiple aspects.
	Above all, event datasets generated from various media differ vastly in structure, syntax and semantics.
	For instance, posts from social networks are often written in spoken language with ungrammatical and fragmented sentences, whereas queries from search engines are usually composed of keywords and highly succinct.
	Secondly, cross-platform analysis has a high requirement on datasets, since the alignment information is usually confidential and sensitive in consideration of user privacy. Although there exists a recent work~\cite{ASNets}, the data and platforms are highly limited.
	Furthermore, the uncertainty and ambiguity of human languages lead to great deal of noise.
	Last but not least, straightforward measures, such as retweet frequencies may not be applicable for all platforms. These methods view all the words or records equally, however, some words or records are not truly associated to the events, so the contribution of different words or records should be different.
	Therefore, it is a meaningful task to analyze cross-platform event popularity.

	In this paper, we intend to explore event popularity between pairwise text media. Our goal is to quantify event popularity by semantic relations, and to provide analytical methods for cross-platform knowledge discovery on large-scale datasets.

	Conceptually, We model the event dissemination trends as \textit{event popularity time series} (EPTS) at any given temporal resolution, encapsulating relevant words and their semantic and lexical relations.
	Inspired by the observation that the diversity of the media and their mutual influences cause the EPTSs to be temporally warped, we seek to identify the alignment between pairwise EPTSs to support deeper analysis.

	Practically, we propose a novel scheme called \textsc{DancingLines} to depict event popularity from pairwise media and quantitatively analyze the popularity trends. After a customized data preprocessing procedure,
	\textsc{DancingLines} facilitates cross-platform event popularity analysis with two innovative models, TF-SW (Term Frequency with Semantic Weight) and $\omega$DTW-CD ($\omega$eighted Dynamic Time Warping with Compound Distance).


	TF-SW is a semantic-aware popularity quantification model based on an integrated weight coefficient that leverages Word2Vec~\cite{ DBLP:conf/nips/MikolovSCCD13} and TextRank~\cite{mihalcea2004textrank}.
	The model includes three steps that the first one is to discard the words unrelated to certain events; then we utilize semantic and lexical relations to get similarity between words and highlight the semantically related ones with a \emph{contributive words} selection process; finally based on similarity, TextRank algorithm gives us the importance of each word, then the popularity of a certain event.
	EPTSs generated by TF-SW are able to capture the popularity trend of a specific event at different temporal resolutions.

	$\omega$DTW-CD is a pairwise EPTSs alignment model using an extended Dynamic Time Warping method.
	It generates sequence of matches between the temporally warped EPTSs. Temporal differences and shape patterns are also considered to avoid the unrealistic \textit{far-match} and \textit{singularity} problem. In addition, we propose three novel
	metrics to evaluate cross-platform EPTSs for various scenarios, providing valuable insights on interpreting event popularity trends.




	
	The main work we do in this paper is to innovatively transform the cross-platform event dissemination analysis, which is hard to quantify, into comparisons, alignments, and interpretations of two event popularity time series.

	Experimental results on eighteen real-world event datasets from Baidu, the most popular search engine in China, and Weibo, Chinese version of Twitter, validate the effectiveness and applicability of our models. We demonstrate that TF-SW
	is in accordance with real trends and sensitive to burst phases, and that $\omega$DTW-CD successfully aligns EPTSs. 
	The model not only gives an excellent performance, but also shows superior robustness.
	Moreover, we also apply \textsc{DancingLines} in different cases to illustrate the insights and advantages of our scheme.
	In all, \textsc{DancingLines} has broad application potentials to reveal knowledge of various aspects of cross-platform events and social media.

	The contributions of this paper are summarized as follows:
	\begin{itemize}
		\item We design a novel scheme called \textsc{DancingLines} with two models to analyze cross-platform event popularity, which quantifies event popularity with semantic and lexical relations, and aligns temporally warped EPTSs by an extended Dynamic Time Warping method.


		\item Three novel metrics are proposed to evaluate event popularity trends, providing valuable insights for cross-platform knowledge discovery. 


		\item We validate the effectiveness and applicability of \textsc{DancingLines} via experiments on eighteen real-world event datasets from a social network and a search engine.

	\end{itemize}

	The rest of this paper is organized as follows. In Section~\ref{sec:relatedwork}, related work is discussed. In Section~\ref{sec:probfor}, we define the problem. In Section~\ref{sec:overview}, we introduce the overview of \textsc{DancingLines}. The two models TF-SW and $\omega$DTW-CD are discussed in details respectively in Section~\ref{sec:model1} and Section~\ref{sec:model2}. Section~\ref{sec:verification} verifies \textsc{DancingLines} and presents case studies on real-world datasets from Weibo and Baidu.
	Finally, we conclude the paper in Section~\ref{sec:conclusion}.



	%

	\section{Related Work}
	\label{sec:relatedwork}
	\textbf{\textit{Event Popularity Analysis.}}
	Many researches~\cite{wikitopics, cikmevent1,DBLP:conf/kdd/LeeLM13, WISE2016/ESAPCroPlatTreAnaSNSE} have focused on event evolution analysis for a single medium. The event popularity was evaluated by hourly page view statistics from Wikipedia in~\cite{wikitopics}. Reference~\cite{DBLP:conf/kdd/LeeLM13} chose the density-based clustering method to group the posts in social text streams into events and tracked the evolution patterns. Breaking news dissemination is studied via network theory based propagation behaviors in~\cite{liu2016breaking}.
	Reference~\cite{WISE2016/ESAPCroPlatTreAnaSNSE} proposed a TF-IDF based approach to analyze event popularity trends. 
	In all, network-based approaches usually have high computational complexity, while frequency-based methods are usually less accurate on reflecting the real event popularity.

	Reference~\cite{DBLP:journals/amc/AbilhoaC14} proposed a keyword extraction method for tweet message collections. However, it missed some valuable semantic information, which could have contributed to better results. Reference~\cite{jin2016labelling} presented a topic labeling model by learning semantic representations of topic words and clustering coherent topic labels. A context-sensitive method based on PageRank to extract topical keyphrases was proposed in \cite{DBLP:conf/acl/ZhaoJHSALL11}. Their work sheds light on integrating lexical, contextual and semantic relations in our TF-SW model.



\textbf{\textit{Cross-Platform Analysis.}}
From a cross-platform perspective, existing researches focus on topic detection, cross-social media user identification, cross-domain information recommendation, etc. Reference~\cite{DBLP:journals/tomccap/BaoXMH15} selected Twitter, \textit{New York Times} and Flickr to represent multimedia streams, and provided an emerging topic detection method. 
An attempt, trying to combine Twitter and Wikipedia to do first story detection, was discussed in~\cite{firststory}.
Reference~\cite{DBLP:journals/tkde/ZhouLZM16} proposed an algorithm based on multiple social networks like Twitter, Weibo, and Facebook to identify anonymous identical users.
The relationship between social trends from social network and web trends from search engine are discussed in \cite{Giummol2013, Kwak2010}. Recently, a good prediction of social links between users from aligned networks using sparse and low rank matrix is well discussed in~\cite{linkpredhan}.
However, few studies have been conducted for event popularity analysis from cross-platform perspective.

	\textbf{\textit{Dynamic Time Warping.}} \textit{Dynamic Time Warping} (DTW) is a well-established method for similarity search between time series.
	Originating from speech pattern recognition~\cite{sakoe1978dynamic}, DTW has been effectively implemented in many domains~\cite{Giummol2013}.
	Recently, remarkable performance on time series classification and clustering by combining KNN classifiers have been achieved in~\cite{maustime,cikmdtw1}.
	Regarding the design of DTW, many efforts have been made to improve the performance and compatibility. The well-known Derivative DTW is proposed in~\cite{keogh2001derivative}. Weighted DTW~\cite{jeong2011weighted} was designed to penalize high phase differences. 
	In \cite{silva2016effect}, the side effect of endpoints which tends to disturb the alignments dramatically in time series is confirmed and an improvement for eliminating such issue is proposed.
	We are inspired by these related works when designing our own DTW based model for aligning EPTSs.

	




	\section{Problem Formulation}
	\label{sec:probfor}
	\subsection{Event Popularity Quantification}
	Our first task is to quantify the popularity of a specific event $\mathscr{E}$. After choosing a certain time resolution, the popularity dissemination trend of the event $\mathscr{E}$ can be represented in the form of \emph{Event Popularity Time Series} on a period of time $T$.





	We start from dividing the time span $T$ of an event into $n$ periods, which is determined by the time resolution, each stamped with $t_i$, $T = \langle t_1,\cdots, t_n\rangle$. A record is a set of words preprocessed from datasets, such as a post from social networks or a query from search engines. The words here can include single words, phrases, number sequences, and other special information related to a certain event, but apart from punctuation, conjunctions, hyper links, etc. Then, we use the notation $w_k^i$ to represent, within time interval $t_i$, the $k$th word in a record. The notation $R^i_j = \{ w^i_1, w^i_2, \cdots , w^i_{|R_j^i|}\}$ is the $j$th record within time interval $t_i$. 


	An \emph{event phase}, corresponded to $t_i$ and denoted as $E_i$, is a finite set of words, and each word is from a related record $R^i_j$. As a result $E_i = \bigcup_jR_j^i$. 

	We can now introduce the prototype of our popularity function $pop(\cdot)$.
	For a given word $w_k^i \in E_i$, the popularity of the word $w_k^i$ is defined as
	\begin{equation} \label{eqn:popularity}
		pop(w_k^i)= fre(w_k^i) \cdot weight(w_k^i),
	\end{equation}
	where $fre(w_k^i)$ is the word frequency of $w_k^i$ within $t_i$. The weight function, $weight(w_k^i)$, for a word within $t_i$, is the kernel we solve in the TF-SW part and is the key to generate event popularity. Recent works on this problem usually regard the frequency of words as the measurement of their popularity, which is not reasonable enough.
	In this work, we propose a weight function not only utilizing the lexical but also taking into account semantic relationships which has better performance and is more reasonable than recent works. 
	Details about how to define the weight function is discussed in Section~\ref{sec:model1}.
	
	Once we get popularity of word $w_k^i$ within $t_i$, the popularity of an event phase $E_i$, $pop(E_i)$, can be generated by summing up all words' popularity,

	\begin{equation} \label{eqn:eventpop}
		pop(E_i)=\sum_{w_i^k \in E_i} pop(w_k^i).
	\end{equation}




	We regard the pair $(t_i, pop(E_i))$ as a data point on X-Y plane
	and get a series of data points, formalizing a curve on the plane to reflect the dissemination trend of an event $\mathscr E$.

To compare the curves from different media, a further normalization is employed,
	\begin{equation} \label{eqn:eventnormal}
		\overline{pop}(E_i)=\frac{pop(E_i)}{\sum\limits_{1 \le k \le n}pop(E_k)}.
	\end{equation}



	After the normalization, the popularity trend of an event on a single medium is represented by a sequence,
	denoted as $\mathscr{E} = \langle \overline{pop}(E_1), \cdots, \overline{pop}(E_n)\rangle$, which is defined as \textit{Event Popularity Time Series}.



	\subsection{Time Series Alignment}
	
\begin{figure}[htb]
		\centering
		\includegraphics[width = \linewidth]{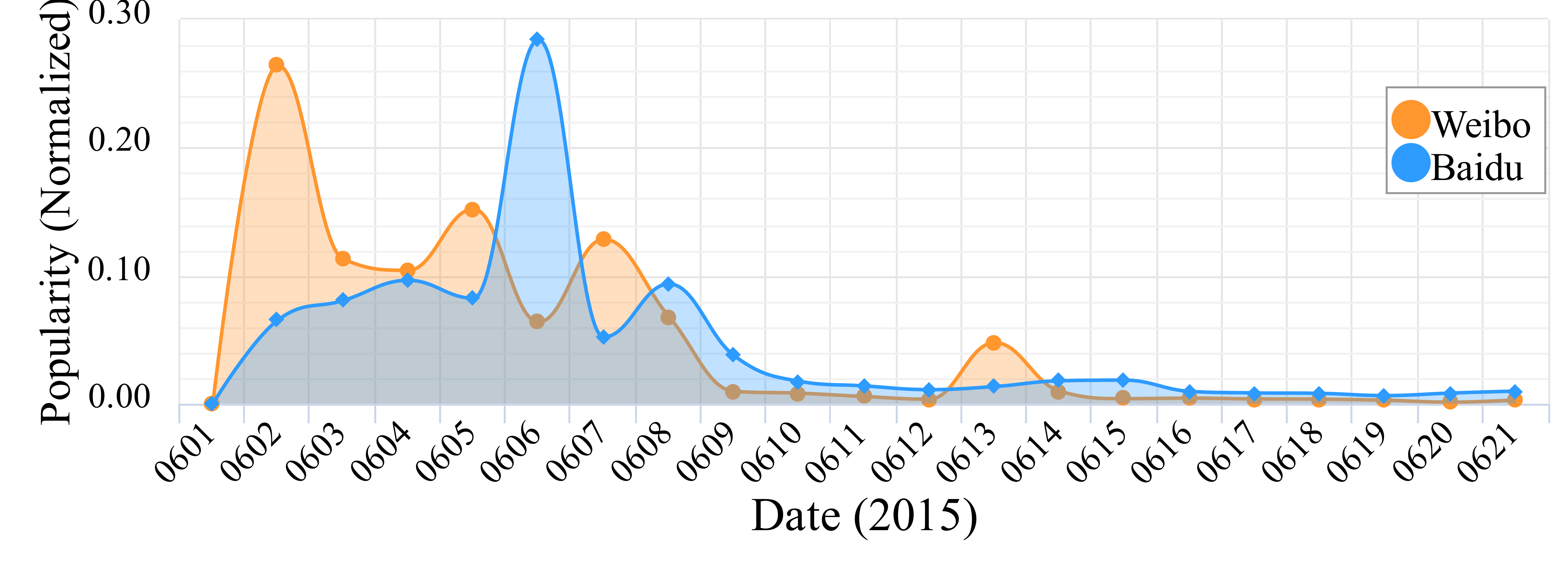}
		\caption{Normalized EPTSs of Event \textit{Sinking of a Cruise Ship}}
		\label{epts}
	\end{figure}

Two EPTSs generated from two platforms of an event $\mathscr{E}$ are now comparable and can be visualized in a same X-Y plane as Fig.~\ref{epts}, which shows normalized EPTSs of Event \textit{Sinking of a Cruise Ship} generated from Baidu and Weibo. A Chinese cruise ship called Dongfang Zhi Xing sank into Yangtze River on the night of June 2, 2015 and the following process lasted for about 20 days. X-axis in Fig.~\ref{epts} represents time and Y-axis indicates the event popularity. Peaks are reached on June 2 and June 6 respectively for Weibo and Baidu. If we shifted the orange EPTS, generated from Weibo, to the right for about 4 units, we would notice the blue one approximately overlaps the orange one. This phenomenon indicates a \emph{temporal warp}, which means the trend features are similar, but there exists time difference between EPTSs.

\begin{figure}[htb]
	\centering
	\includegraphics[width=\linewidth]{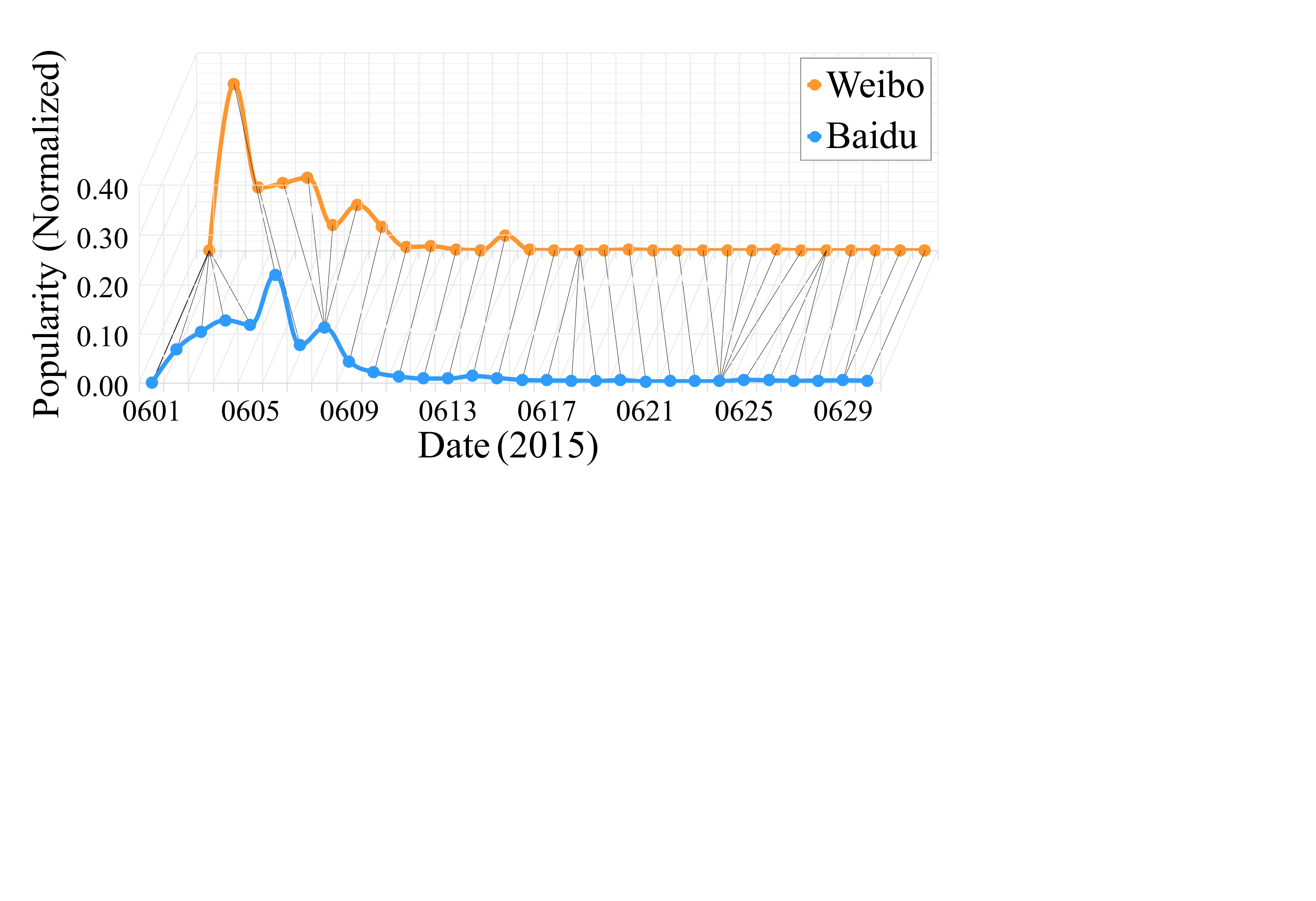}
	\caption{Aligned EPTSs of Event \textit{Sinking of a Cruise Ship}}
	\label{fig:aligndfzx}
\end{figure}

According to Fig.~\ref{epts}, EPTSs are temporally warped due to the special features of events and media. For example, entertainment news tends to be disseminated on social networks and can easily draw extensive attention, but its dissemination on serious media like \textit{Wall Street Journal} is very limited. Another interesting feature is the time difference between EPTSs, namely the degree of temporal warp, which reveals whether events have preference to media. Alignments of EPTSs are quite suitable to reveal such interesting features. Fig.~\ref{fig:aligndfzx} is a demo of an alignment between EPTSs mentioned above.

A link in Fig.~\ref{fig:aligndfzx} represents a match and we can eliminate  time factors through an alignment. A proper alignment between pairwise EPTSs would be helpful to interpret the event popularity and reveal the latent properties of events and media.

\begin{figure*}[htb]
	\centering
	\includegraphics [width=\textwidth]{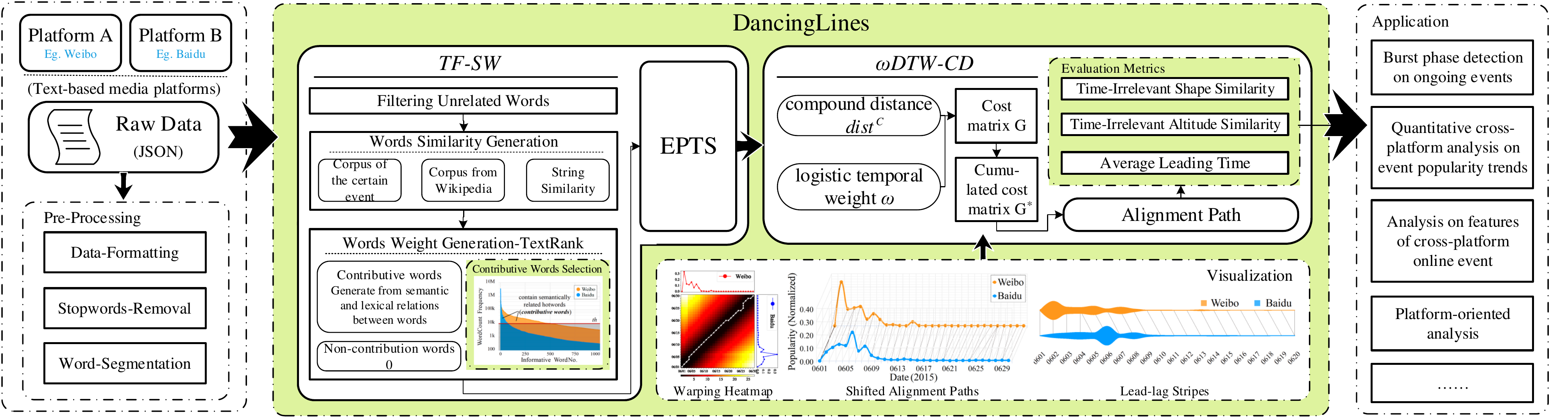}
	\caption{The overview of \textsc{DancingLines} Scheme
	}
	\label{fig:scheme}
\end{figure*}



	Two temporally-warped EPTSs of an event $\mathscr{E}$ from two media $A$ and $B$, are denoted as
	\[\mathscr{E}^* = \langle \overline{pop}(E^*_1)\cdots, \overline{pop}(E^*_n)\rangle,\]
	where $\mathscr{E}^*$ represents either $\mathscr E^A$ or $\mathscr E^B$.

	A \emph{match} $m_k$ between $E^A_i$ and $E^B_j$ is defined as $m_k=(i, j)$. Distance between two matched data points is denoted as $dist(m_k)$ or $dist(i, j)$.


	There is one problem, \emph{twist}, existing when there are two matches $m_{k_1} = (i_1,j_1)$, $m_{k_2} = (i_2,j_2)$ with $i_1 < i_2$, but $j_1 > j_2$.
	The reason why there cannot be \emph{twist} is that time sequence and the evolution of events cannot be reversed. 
	
	The goal of EPTS alignment is to find a series of twist-free matches $M=\{m_1, \cdots, m_{|M|}\}$ for two EPTSs $\mathscr{E}^A$ and $\mathscr{E}^B$ that every data point from an EPTS has at least one counterpoint from the other one, and the cumulative distance is the minimum. An intuitive thinking about an optimal alignment is that it should be a feature-to-feature one and differences between aligned EPTSs should be as small as possible. The minimum cumulative distance satisfy these two requirements. The key of alignments is to define a specific, precise, and meaningful distance function $dist(\cdot)$ for our task, which will be fully discussed in Section~\ref{sec:DTW}.

	We describe the problem,
	as a linear programming problem,
	\begin{gather}
\min\sum\limits_{k=1}^{|M|} dist(m_k),\label{eqn:minimum} \\
\{E_i^* \mid \forall m_k,\ E_i^*\not\in m_k\} = \emptyset,\ \forall 1 \le i \le n,\label{eqn:counter}\\
\forall m_{k_1}=(i_1,j_1),\ \forall m_{k_2}=(i_2,j_2),\ if\ i_1 \leq i_2,\ j_1\leq j_2.\label{eqn:twist}
	\end{gather}
	
	The target is to get the minimum cumulated distance guaranteed by Eqn.~\eqref{eqn:minimum}.
The other two conditions for optimal alignments are that each data point should have a counterpoint from the other EPTS, or in another word, every data point should appear in at least one match $m_k$, ensured by Eqn.~\eqref{eqn:counter}
and that alignment should be twist-free, showed in Eqn.~\eqref{eqn:twist}.

	\section{Scheme Overview of \textsc{DancingLines}}
	\label{sec:overview}



In this paper, we design an innovative scheme, \textsc{DancingLines}, which can precisely capture and quantify the event popularity between pairwise media and then analyze the event trend disseminations. The overview of \textsc{DancingLines} is illustrated in Fig.~\ref{fig:scheme}. We first preprocess the data,
	then implement the TF-SW and $\omega$DTW-CD models, and finally apply our scheme to real event datasets.

	\textbf{Data Preprocessing} is applied on the raw data, which is organized in json format, extracted from two different platforms  with three steps. First of all, in Data-Formatting step,  we filter out all irrelevant characters, such as punctuation, hyper links, etc. Secondly, Stopword-Removal step cleans frequently used conjunctions, pronouns and prepositions. Finally, we split every record into words utilizing an existing open-source word segmentation tool through Word-Segmentation step.

	\textbf{TF-SW} is a semantic-aware popularity quantification model based on Term Frequency with Semantic Weight, which leverages Word2Vec and TextRank to generate EPTSs at certain temporal resolutions. This model is established by three steps. First of all, a cut-off mechanism is proposed to filter the unrelated words. Secondly, we construct TextRank graph to calculate the relative importance for the remaining words. Finally, a synthesized similarity calculation with semantic and lexical relations is defined for the edge weights in TextRank graph. In addition, we find that only the words with both high semantic and lexical relations with other ones truly determine the event popularity. For that, a conception \emph{contributive words} is defined and will be discussed in Section~\ref{sec:model1}.





	\textbf{$\omega$DTW-CD} is a pairwise EPTSs alignment model derived from the Dynamic Time Warping method with Compound Distance. In this model, we innovatively define three distance function for DTW, event phase distance $dist^\mathscr{E}(\cdot)$, derivative distance $dist^D\left(\cdot\right)$, and Euclidean vertical line distance $dist^L\left(\cdot\right)$. Based on these three distance function, a compound distance is generated. A temporal weight coefficient is also introduced into the model and play an important role in improving the alignment results. Moreover, three metrics are proposed to evaluate event popularity time series and their alignments. We further introduce these in detail in Section~\ref{sec:model2}.

The whole scheme has a lots of potential applications, for example, it can be used to do first story detection, analysis on event dissemination trends, evaluation of online platforms, etc. In this paper, we focus mainly on how and why our scheme works and have better performance than other methods. The potential applications are also covered in case studies part.



	\section{Semantic-Aware Popularity Quantification Model (TF-SW)}	\label{sec:model1}
In this section, we mainly discuss the definition of weight function $ weight\left(\cdot\right) $ in Eqn.~\eqref{eqn:popularity}. The problem is that how do we evaluate the weight, or in another word, relative importance of every word in a huge corpus. A similar problem faced by the researchers of search engines is to rank the Web pages mechanically and efficiently according to the queries. A well-developed method PageRank, proposed to solve this problem, inspires us and we adopt TextRank~\cite{mihalcea2004textrank}, an adaption from PageRank, to address this weight evaluation problem. TextRank constructs an undirected weighted graph, making an extension by giving weights to edges, compared to the original PageRank. To calculate weights of edges, we adopt Word2Vec to quantify the relationship between any pair of words.

This section includes four parts that the first one is about filtering unrelated words, second one is the construction of TextRank graph, the third part discusses the quantification of relationships between words, and finally we give the definition of weight function $weight\left(\cdot\right)$.


\subsection{Filtering Unrelated Words}\label{sec:filter}
Since the number of distinct words for an event can be thousands of hundreds and there are tons of ones that are actually not related to the event at all, it is too expensive to take them all into account. We propose a cut-off threshold mechanism to eliminate these unrelated noisy words and significantly reduce the complexity of whole scheme.

\begin{figure}[htb]
		\centering
		\includegraphics[width=\linewidth]{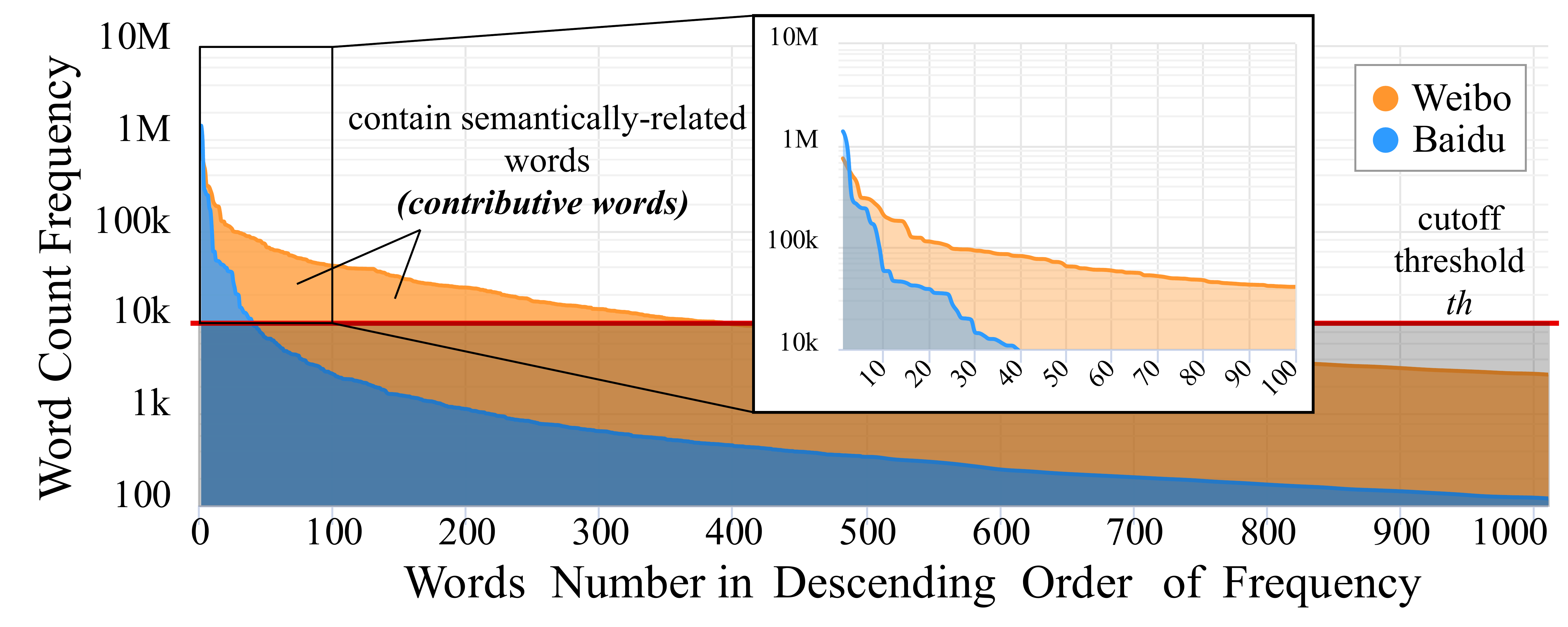}
		\caption{Word frequencies of Event \textit{AlphaGo}}
		\label{fig:wordcntfre}
	\end{figure}

To illustrate, in Fig.~\ref{fig:wordcntfre}, we count every distinct words' frequency of Event \textit{AlphaGo} whose 4-1 victory against legendary Go player Mr Lee Sedol in Seoul, South Korea, in March 2016 was watched by over 200 million people worldwide. When observing the word frequency distribution, we find a long tail phenomenon.
In fact, datasets like natural language corpus approximately obey the power law distribution
and Zipf's Law~\cite{zipf}. Denoting $r$ as the frequency rank of a word in a corpus and $f$ as the corresponded word's frequency, then
	\begin{equation}
f = H\cdot r^{-\alpha},
		\label{zipf}
	\end{equation}
where $\alpha$ and $H$ are feature parameters for a specific corpus which can be obtained through linear regression.

Since the words with high frequency is the necessary but not sufficient condition for those words to really reflect the actual event trends, an interesting question that where the majority of distribution of $r$ lies is raised. For any power law with exponent $\alpha>1$, the median is well defined~\cite{zipf}. That is, there is a point $r_{1/2}$ that divides the distribution in half so that half the measured values of $r$ lie above $r_{1/2}$ and half lie below. In our case, $r$ as rank, its minimum is $1$, and the point is given by
\[\int_{r_{1/2}}^\infty f\ dr = \frac{1}{2}\int_{r_{min}}^\infty f\ dr\]
or
\[r_{1/2}=2^{1/\left(\alpha - 1\right)}r_{min}=2^{1/\left(\alpha - 1\right)}.\]

Emphasis should be placed on the words that rank ahead of $r_{1/2}$, and the words within the long tail which are occupied by noise should be discarded. Thus cut-off threshold can now be defined as
\begin{equation}
th = H\cdot r_{1/2}^{-\alpha}=\frac{1}{2}\cdot H\cdot 2^{1/\left(1-\alpha\right)}
\label{zipfth}
\end{equation}

Through this filter, we dramatically reduce the whole complexity of the scheme. As illustrated in the subfigure of Fig.~\ref{fig:wordcntfre}, for Event \textit{AlphaGO}, the words we need to consider for Baidu reduce from thousands to around 40 and the ones for Weibo reduce to about 350, so the  complexity has been reduced by at least 3 orders of magnitude, saving lots of time and calculation, which is appreciated in the fields of big data mining.


%

\subsection{Construction of TextRank Graph}
After filtered through threshold, the remaining words are regarded as the representative words that do matter in quantifying the event popularity.

However, for the remaining words, the importances are still obscure. They cannot just be naively presented by words' frequency, as a result we introduce TextRank~\cite{mihalcea2004textrank} into our scheme to solve the definition of the weight function $weight(\cdot)$. 

The basic idea of TextRank is derived from the classical PageRank algorithm, which is calculated on a directed unweighted graph that each vertex is a Web page and each directed edge represents a hyper link. For our task here, vertex stands for a word that has survived the frequency filter in Section~\ref{sec:filter} and we use undirected edges in TextRank instead of directed edges in PageRank, since the relationships between words are bidirectional.

Inspired by the idea of TextRank, we further need to define the weights of edges in the graph described above. The weights of edges are supposed to reflect the relative relationships between two words that are connected, so we introduce a conception \emph{similarity} between words $w_i$ and $w_j$, denoted as $sim\left(w_i,w_j\right)$ to represent these relations.

However, we notice that there exist some words which passed the first filter but having negative similarity with all the other remaining words, which means these words are semantically far away from the topic of events. This phenomenon, in fact, indicates the existence of paid posters who post a large number of unrelated messages especially on social networks. To address this problem, we focus on the really related words and define a conception \emph{contributive words}, denoted as
\begin{equation}\label{eqn:contributive}
	C_i = \{w^i_j \in E_i \mid \exists w^i_k \in E_i, \ sim(w^i_k, w^j_k)>0\}
\end{equation}
 and $\mathscr{C} =\bigcup
C_i$.
It is worth pointing out that this another filter-like process does not increase any computational complexity and we just do not establish edges when their weights are less than zero, then the non-contributive words will be discarded.

We construct a graph for each event phase $E_i$, where vertices represent the words and edges refer to their similarity $sim(w_i, w_j)$. We run the TextRank algorithm on the graphs and then get the real importance of each contributive word, $ TR(w_i) $. The formula for TextRank is defined as
\[TR(w_i) = \frac{1-\theta}{\left|\mathscr{C}\right|} + \theta \cdot \sum\limits_{j\rightarrow i}\frac{sim(w_i,w_j)}{\sum\limits_{k\rightarrow j}sim(w_k, w_j)} \cdot TR(w_j),\]
where the factor $\theta$, ranging from $ 0 $ to $ 1 $, is the probability to continue to random surf follow the edges, since the graph cannot be a perfect graph and face potential dead-ends and spider-straps problem
in practice. According to \cite{mihalcea2004textrank}, $\theta$ is usually set to be 0.85. $\left| \mathscr{C}\right|$ represents the number of all contributive words, and $j\rightarrow i$ refer to words that is adjacent to word $w_i$.

\subsection{Similarity Between Words}
In our view, similarity between words are contributed by their semantical and lexical relationships and these two parts will be discussed in this subsection.

First of all, to quantify words' semantic relationships, we adopt Word2Vec~\cite{DBLP:conf/nips/MikolovSCCD13} to map word $w_k$ to vector $\mathbf{w}_k$ in a high-dimensional space. To comprehensively reflect the event characteristics, we integrate two corpora, an event corpus $\mathbb{R}$ from our datasets and a supplementary corpus extracted from Wikipedia with a broad coverage of events (denoted as \emph{Wikipedia Dump}, or $\mathbb{D}$ for short), to train our Word2Vec models.
For a word $w_k$, the corresponding word vectors are $\mathbf{w}^\mathbb{R}_k$ and $\mathbf{w}^{\mathbb{D}}_k$ respectively. Both event-specific and general semantic relations between words $w_i$ and $w_j$ are extracted and composed by
\[sem(w_i, w_j) = \beta\cdot\frac{\mathbf{w}^\mathbb{R}_i \cdot \mathbf{w}^\mathbb{R}_j}{\lVert\mathbf{w}^\mathbb{R}_i\rVert \cdot \lVert\mathbf{w}^\mathbb{R}_j \rVert} + \left(1-\beta\right) \cdot \frac{\mathbf{w}^\mathbb{D}_i\cdot \mathbf{w}^\mathbb{D}_j}{\lVert\mathbf{w}^\mathbb{D}_i\rVert \cdot \lVert\mathbf{w}^\mathbb{D}_j\rVert},\]
where $\beta$ is related to the two corpora and determines which one and to what extent we would like to emphasize.

Secondly,
we consider the lexical information and integrate the string similarity so that we can combine the
\[sim(w_i, w_j) = \gamma\cdot sem(w_i, w_j) + \left(1-\gamma\right) \cdot str(w_i, w_j),\]
where we introduce a parameter $\gamma$
to make our model flexible and general to different languages. For example, those words that look similar are likely to be related in English, while this likelihood is fairly limited for languages like Chinese.
We adopt the
efficient cosine string similarity as
\[str(w_i, w_j)=\frac{\sum\limits_{c_l \in w_i \cap w_j} num(c_l, w_i) \cdot num(c_l, w_j)}{\sqrt{\sum\limits_{c_l \in w_i} num(c_l,w_i)^2}\cdot\sqrt{\sum\limits_{c_l \in w_j}\!num(c_l,w_j)^2}},\]
where $num(c_l, w_i)$ means counts of character $c_l$ in word $w_i$.

\subsection{Definition of Weight Function}
Since the sum of vertices' TextRank values for a graph is always $1$ regardless of the graph scale, the TextRank value tends to be lower when there are more contributive words within the time interval. Therefore, a compensation factor within each event phase $E_i$ is multiplied to the TextRank values, and the weight function $weight(\cdot)$ for contributive words is finally defined as
\[weight(w^i_j) = \frac{TR(w^i_j)}{\left|C_i\right|}\cdot\sum\limits_{w^i_k \in E_i}fre(w^i_k).\]

Recalling that in our scheme, the event popularity $pop(E_i)$ is the sum of popularity of \emph{all} words, for the consistency of Eqn.~\eqref{eqn:popularity}, we make the weight function for the non-contributive words identically equal to zero.
Then for all words, popularity can be calculated through Eqn.~\eqref{eqn:popularity}. For each event phase $E_i$, according to Eqn.~\eqref{eqn:eventpop}, we can generate the event popularity within $t_i$ and EPTSs through Eqn.\eqref{eqn:eventnormal}.



\begin{figure}[ht]
\centering
\includegraphics[width=\linewidth]{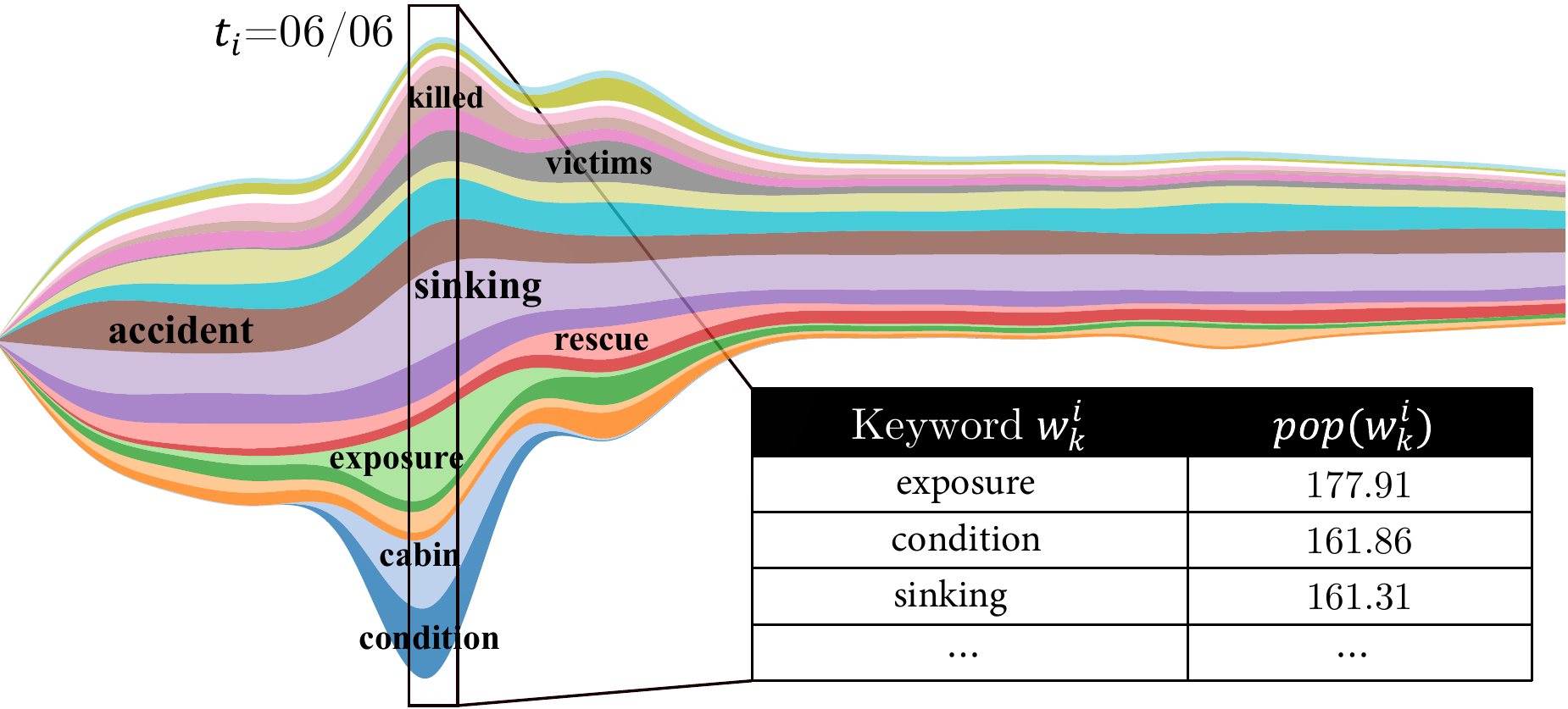}
\caption{Word Stream  of Event \textit{Sinking of a Cruise Ship} on Baidu with an event phase highlighted}
\label{wordstream}
\end{figure}

To describe the way how contributive words contribute to the event popularity on a single medium, we visualize a demo, implemented by weighted wiggle method~\cite{DBLP:journals/tvcg/ByronW08}, in Fig.~\ref{wordstream}. It illustrates how the event popularity is composed by the popularity of contributive words. Each colored stripe corresponds to a contributive word and its width indicates the relative word's popularity. We can see that the Event \textit{Sinking of a Cruise Ship}'s popularity reached its peak on June 6, 2015 and in that day, the stripe related to the word \emph{exposure} occupies the biggest width, which indicates this word contributes the most significant part to that day's event popularity. Similar examples are \emph{condition}, \emph{sinking}, \emph{accident}, \emph{rescue}, etc. In addition, some of contributive words' popularity in June 6 are listed in the table from Fig~\ref{wordstream}.

%


\section{Cross-platform Analysis Model ($\omega$DTW-CD)}\label{sec:model2}


To interpret the temporal warp between EPTSs, a time series alignment method mentioned in Section~\ref{sec:probfor} is proposed.
In this section, we first discuss the classic DTW method with Euclidean distance and show how the optimal alignment is generated. Next, we integrate the DTW~\cite{sakoe1978dynamic} method with an innovative distance function, event phase distance, to solve the similar problem. Then, we talk about compound distance function in the third part. Finally, to evaluate how well an alignment method does and interpret the alignments, we propose three metrics for evaluating time series alignments.

\subsection{Classic Dynamic Time Warping with Euclidean Distance}
\label{sec:classicDTW}

Suppose two time series $\mathscr E^A = \{E^A_1, E^A_2, \cdots, E^A_{\left|\mathscr E^A\right|}\}$ and $\mathscr E^B = \{E^B_1, E^B_2, \cdots, E^B_{\left|\mathscr E^B\right|}\}$, which are temporally warped but not aligned yet. The goal of DTW is to find an optimal warping path to align the two sequences. Classic DTW starts by constructing a cost matrix $G_{\left|\mathscr E^A\right|\times \left|\mathscr E^B\right|}$, where each $g_{i,j}$ represents a distance $ dist\left(i,j\right)$ between $E_i^A$ and $E_j^B$.
A cumulated cost matrix $G^*_{\left|\mathscr E^A\right|\times \left|\mathscr E^B\right|}$ is obtained by calculating the minimum sum of distances $dist^*\left(i,j\right)$ iteratively from $dist\left(1,1\right)$ and we denote it as $d_{i,j}$ in the following two equations for the limitation of space.
\begin{equation}
d^*_{i,j} =
\begin{cases}
\sum_{k=1}^{i} d_{k,j}, &
j=1 \\
\sum_{k=1}^{j} d_{i,k}, & i=1
 \\
d_{i,j} + \min\{d^*_{i-1,j}, d^*_{i-1,j-1}, d^*_{i,j-1}\}, & else.
\end{cases}
\label{eq:accugen}
\end{equation}

Then, by going through cost matrix $G^*_{\left|\mathscr E^A\right|\times \left|\mathscr E^B\right|}$ backwards from $dist^*\left(\left|\mathscr E^A\right|,\left|\mathscr E^B\right|\right)$, the optimal alignment $M$, a sequence of matches between $E_i^A$ and $E_j^B$, is generated by
\begin{equation}
m_k =
\begin{cases}
(i,j-1), &i=1 \\
(i-1,j), &j=1 \\
\arg\min\{d_{i-1,j-1},d_{i-1,j},d_{i,j-1}\}, & else.
\end{cases}
\label{eq:alipagen}
\end{equation}


However, we find that, with only the global minimum cost considered, classic DTW with Euclidean distance may provide results suffering from \textit{far-match} and \textit{singularity} problems when aligning pairwise cross-platform EPTSs.

\textbf{Far-Match Problem:} Classic DTW disregards the temporal range, which may lead to \textit{``far-match''} alignments. Since the EPTSs of an event from different platforms keep pace with the event's real-world evolution, alignment of EPTSs' data points that are temporally far away is against the reality.
Thus, classic method should be more robust and Euclidean Distance is not ideal enough for EPTS alignment.

\textbf{Singularity Problem:} Classic DTW with Euclidean distance is vulnerable to the \textit{``singularity''} problem elaborated in~\cite{keogh2001derivative}, where a single point in one EPTS is unnecessarily aligned to multiple points in another EPTS.
%
These singular points will generate misleading results for further analysis.

\subsection{Event Phase Distance}

We already talked about how to align two EPTSs, but we just take the Euclidean distance function for granted and face two fatal limitations. In this subsection, a specific distance, \emph{event phase distance}, will be initially defined.

Recalling Eqn.~\eqref{eqn:contributive} that all the contributive words for an event phase $E_i$ are denoted as $C_i$ and $\mathscr{C}$ is a set of all contributive words for an event $\mathscr{E}$ on single medium, we can utilize the similarity between the contributive word sets $C_i$, obtained from two platforms, to match those event phases.
To quantify this similarity, we propose our \textit{event phase distance} measure. Distance between $E_i^A$ and $E_j^B$ is denoted as $dist^\mathscr{E}\left(i, j\right)$.

Since $\mathscr{C}$ for different platforms are probably not identical, let the general $\mathscr{C}'=\mathscr{C}^A\cup\mathscr{C}^B$, where $\mathscr{C}^A,\mathscr{C}^B$ represent the contributive word sets generated from two platforms.

Then, each word list $C_i$ can be intuitively represented as a one-hot vector $\mathbf{z}_i\in\{0,1\}^{\left|\mathscr{C}'\right|}$, where each entry of vectors indicates whether corresponding contributive word exists in word list $C_i$. However, problem arises when calculating the similarity between these very sparse vectors, especially when the event corpus is of a large scale and there are huge amount of data points in EPTSs, which would increase the complexity exponentially. To address this problem, we leverage \textsc{SimHash}~\cite{simhash02}, adapted from \textit{locality sensitive hashing} (LSH)~\cite{LSH98}, to hash the very sparse vectors to small signatures while preserving the similarity among the words.

According to~\cite{simhash02}, $s$ projection vectors $\mathbf{r}_1,\mathbf{r}_2,\cdots,\mathbf{r}_s
$ are selected at random from the $\left|\mathscr{C'}\right|$-dimensional Gaussian distribution. The LSH function $h_{\mathbf{r}_l}(\mathbf{z}_i)$ is defined as follows:
\begin{equation}
h_{\mathbf{r}_l}(\mathbf{z}_i) = \left\{ \begin{array}{cc}
+1 & \text{if } \mathbf{z}_i\cdot\mathbf{r}_l\geq 0\\
-1 & \text{if } \mathbf{z}_i\cdot\mathbf{r}_l < 0
\end{array}
\right.
\end{equation}

A projection vector $\mathbf{r}_l$ is actually a hash function that hashes a one-hot vector $\mathbf{z}_i$ generated from $C_i$ to a scalar $ -1 $ or $ 1 $. $s$ projection vectors hash the original sparse vector $\mathbf{z}_i$ to a small signature $\mathbf{e}_i$, where $\mathbf{e}_i$ is an $s$-dimensional vectors with entries equal to $ -1 $ or $ 1 $. Sparse vectors $\mathbf{z}_i^A$ and $\mathbf{z}_j^B$ corresponded to data points $E_i^A$ and $E_j^B$ can be hashed to $\mathbf{e}^A_i$ and $\mathbf{e}^B_j$ and the distance between these two data points can be calculated by
\begin{equation}
dist^\mathscr{E}\left(i, j\right) = 1- \frac{\mathbf{e}^A_i\cdot \mathbf{e}^B_j}{\lVert\mathbf{e}^A_i\rVert\cdot\lVert\mathbf{e}^B_j\rVert}.
\end{equation}

The dimension of short signatures, $s$, can be used to tune the accuracy we want to remain versus the low complexity. If we want to dig some subtle information in a high temporal resolution, say half an hour, we should increase $s$ to get more accuracy, while if we just want to have a glimpse of the event, a small $s$ is reasonable.

\subsection{The $\omega$DTW-CD model} \label{sec:DTW}

\begin{figure*}[htb]
	\begin{subfigure}[b]{0.16\linewidth}
		\centering
		\includegraphics[width=\linewidth]{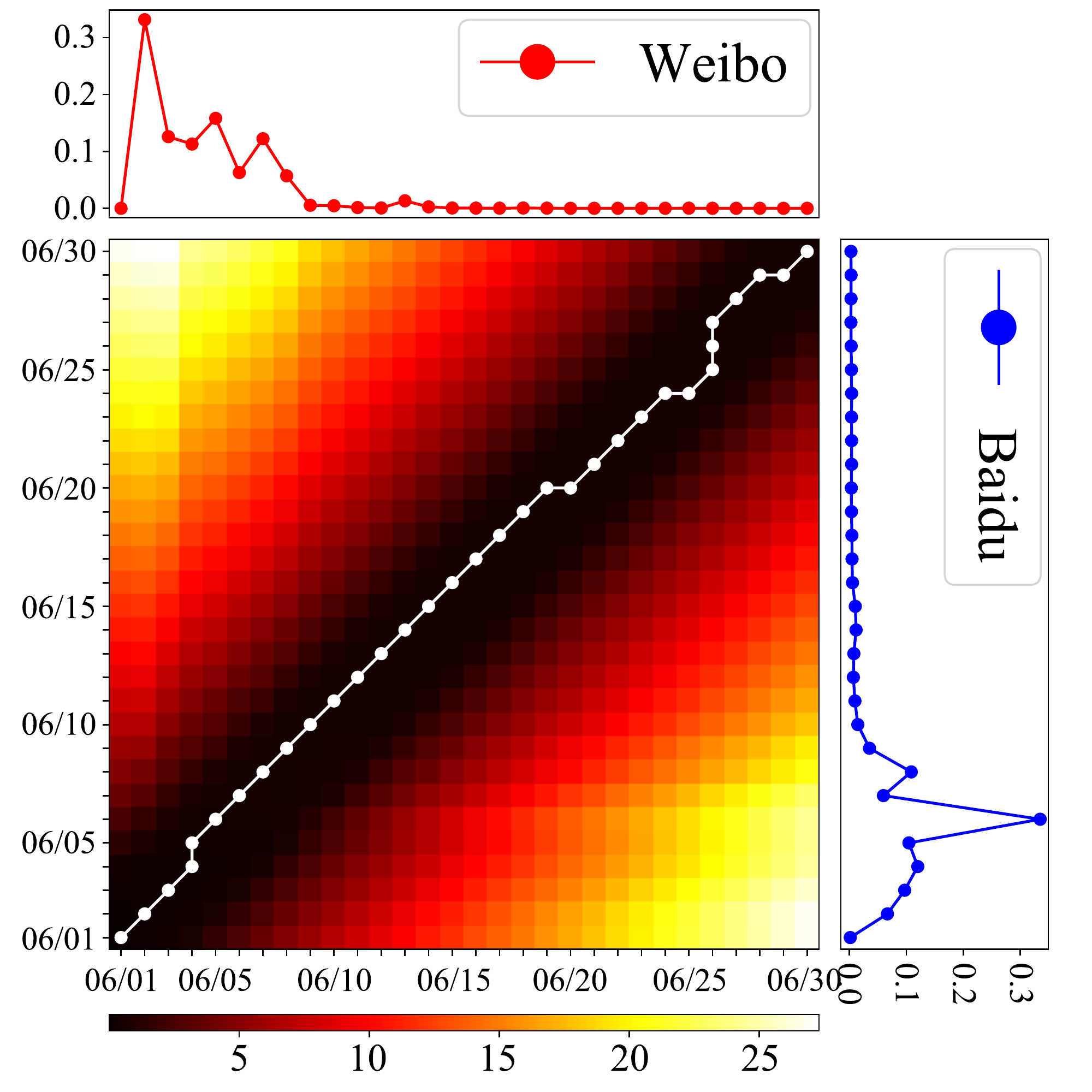}
		\caption{DTW heatmap}
		\label{fig:heatmap}
	\end{subfigure}%
	\begin{subfigure}[b]{0.35\linewidth}
		\centering
		\includegraphics [width=0.96\linewidth]{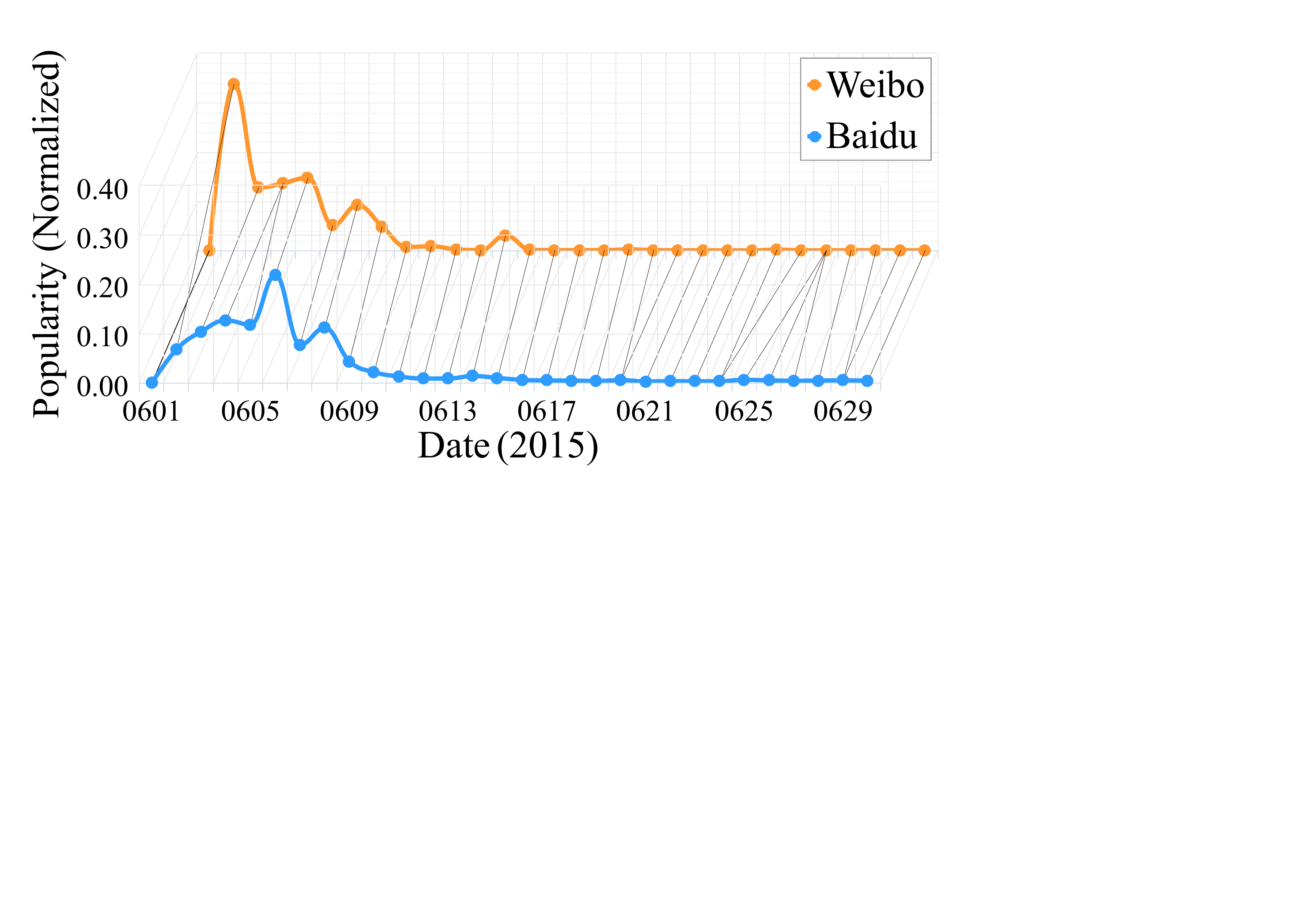}
		\caption {Aligned EPTSs}
		\label{fig:3dalign}
	\end{subfigure}%
	\begin{subfigure}[b]{0.46\linewidth}
		\centering
		\includegraphics [width=\linewidth]{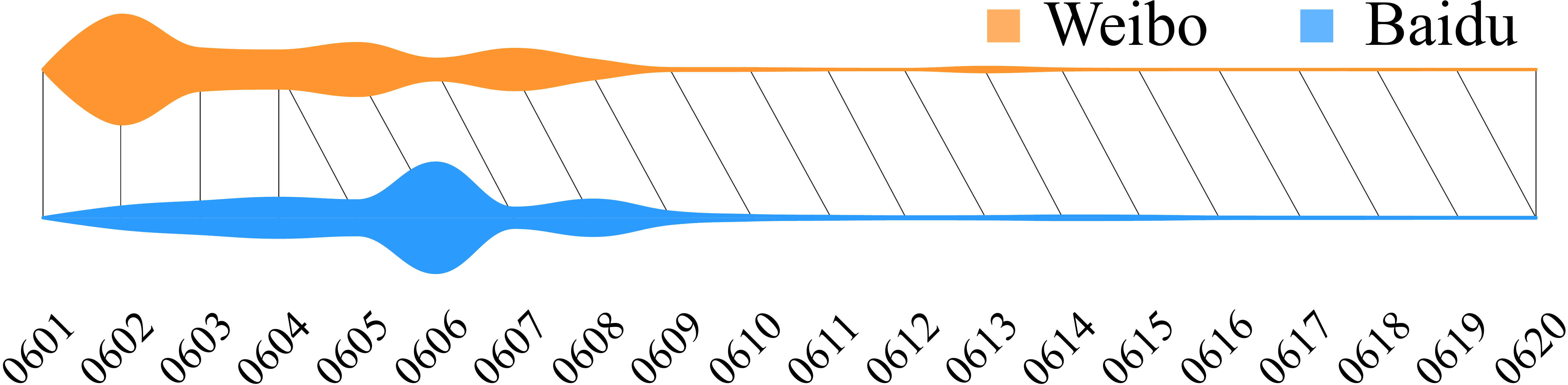}
		\caption {Lead-Lag stripes for aligned EPTSs 
}
		\label{fig:leadlag}
	\end{subfigure}%
	\caption{Visualized results of $\omega$DTW-CD on Event \textit{Sinking of a Cruise Ship} from Baidu and Weibo}
	\label{fig_wdtwcdvis}
\end{figure*}

%
%
%

To more comprehensively measure the distance between data points from two EPTSs, a $\omega$eighted DTW method with Compound Distance ($\omega$DTW-CD) is proposed to balance temporal alignment and shape-matching. $\omega$DTW-CD tries to synthesize trend characters,
Euclidean vertical line distance, and event phase distance all together and this overall distance is measured by compound distance $dist^C\left(i,j\right)$,
\begin{equation}
	dist\left(i,j\right) = dist^C\left(i,j\right) + \omega_{i,j}.
	\label{eq:weightedcost}
\end{equation}

We consider the difference between estimated derivative of the EPTS data points, $dist^D\left(i, j\right)$, as the trend characters distance. According to~\cite{keogh2001derivative}, $dist^D\left(i, j\right)$ generated by
\begin{equation}\label{eqn:derivative}
dist^D\left(i, j\right) = \left| D(E_i^A) - D(E_j^B) \right|,
\end{equation}
where the estimated derivative $D(x)$ is calculated through
\begin{equation}
D(x) = \frac{x_i-x_{i-1}+\frac{x_{i+1}-x_{i-1}}{2}}{2}.
\label{eq:deriv}
\end{equation}
As stated in~\cite{keogh2001derivative}, this estimate is simple but robust to trend characters compared to other estimation methods.

The compound distance $dist^C\left(i, j\right)$ is generated by
\begin{equation}
dist^C\left(i, j\right) = \sqrt[3]{dist^\mathscr{E}\left(i, j\right) \cdot dist^L\left(i, j\right) \cdot dist^D\left(i, j\right)},
\label{eq:costnew}
\end{equation}
where $dist^\mathscr{E}\left(i, j\right)$ is the event phase distance and $dist^L\left(i, j\right)$ is the Euclidean vertical line distance between data points $E^A_i$, $E^B_j$ defined as $dist^L\left(i, j\right) = |E_i^A-E_j^B|$.



In addition, 
for the purpose of flexibility~\cite{jeong2011weighted}, we introduce a sigmoid-like temporal weight
\begin{equation}
\omega_{i,j} = \frac{1}{1+ e^{-\eta(|i-j|-\tau)}}.
\label{eq:temweight}
\end{equation}

The temporal weight is actually a special cost function, similar to the cost function in Machine Learning fields, for the alignment in our task. It has two parameters, $\eta$ and $\tau$, to generalize for many other events and languages. Parameter $\eta$ decides the overall penalty level, which we can tune for different EPTSs. Factor $\tau$ is a prior estimated time difference, having the same unit as the temporal resolution we choose, between two platforms based on the natures of different medias. For example, if breaking entertainment news always disseminate on social networks 2 hours, at most, ahead of search engines, then we set $\tau=2$ hours. If $|i-j|$ is less than 2, it leads to a relative low penalty. While if $|i-j|$ is greater than 2, it will suffer a high penalty, which is quite reasonable since the data points, having large time difference, cannot be aligned in practice. The introduction of sigmoid-like temporal weight significantly reduces the \textit{far-match} problem.


With the weighted compound distance function, the cumulated cost matrix $G^*_{\left|\mathscr{E}^A\right|\times \left|\mathscr{E}^B\right|}$ and the alignment path, namely twist-free matches $M$, can be generated through Eqn.~\eqref{eq:accugen} and Eqn.~\eqref{eq:alipagen} respectively. We visualize the matches in a DTW heatmap in Fig.~\ref{fig:heatmap}, where the white lines indicates the paths and it is not always from the bottom left to the upper right corner and this is the temporal warp we discussed earlier. Another visualization is showed in Fig.~\ref{fig:3dalign} and it gives a direct way to know how the data points from EPTSs are aligned. The links in the figure represent matches.
The lead-lag stripes~\cite{zhong2016tracking} in Fig.~\ref{fig:leadlag} show a more obvious way to know matches. The X-axis represents time and the stripes' vertical width indicates the event popularity in that day. We can find that after the Event \textit{Sinking of a Cruise Ship} happens, the Weibo platform captured and propagated the topic faster than Baidu did in the beginning and then more people started to search on the Baidu for more information so the popularity on Baidu rose.

\subsection{The Design of Three Metrics}
\label{sec:criteria}

To evaluate whether a method does a good job in time series alignments and explore further applications as well as interpretations about $\omega$DTW-CD, we propose three metrics to evaluate and analyze the aligned cross-platform EPTSs. In this section, we only consider EPTSs with the same time span.

\subsubsection{Time-Irrelevant Shape Similarity $\psi_S$}

The time-irrelevant shape similarity $\psi_S$ is a metric measuring the similarity of the shape patterns between the aligned pairs of data points.
It calculates the average sum of the estimated derivative distances $dist^D$ of matches $m_k$ in $M$.
\begin{equation}
\psi_S = 1- \frac{1}{\left| M\right|} \cdot \sum_{m_k\in M} dist^D\left(m_k\right).
\end{equation}

$\psi_S$ indicates the difference between popularity trends of two aligned EPTSs without regard for time difference. This metric is just aimed at testing whether an event have the similar dissemination trends on pairwise media.
Thus, a lower $\psi_S$ may be generated from two weakly coupled media for example, Wikipedia and Twitter, or from an event with different preferences to different media.

\subsubsection{Time-Irrelevant Altitute Similarity $\psi_A$}

The time-irrelevant altitude similarity $\psi_A$ is a metric measuring the similarity of altitudes between the aligned data points from the EPTSs.
It calculates the average sum of the Euclidean vertical line distances $dist^L$ of matches $m_k$ in $M$.
\begin{equation}
	\psi_A = 1- \frac{1}{\left| M\right|} \cdot \sum_{m_k\in M} dist^L\left(m_k\right).
\end{equation}

$\psi_A$ indicates the degree of difference between two aligned EPTSs without regard for time difference. Thus, a lower $\psi_A$ indicates higher sensitivity of the event to different media. For instance, if an event was extremely popular on one platform but not on the other one, $\psi_A$ value would be low.

\begin{table*}[htb]
	\caption{Overall information of the datasets}
	\label{datasetsinfo}
	\centering
	\begin{threeparttable}
		\centering
		\begin{tabular}{ccccccccc}
			\hline
			\multirow{2}{*}{No.} & \multirow{2}{*}{Event Name} &
			\multirow{2}{*}{Category} & \multirow{2}{*}{Date} & \# of & \multicolumn{2}{c}{\# of Records ($k$)} & \multicolumn{2}{c}{Size (MB)} \\\cline{6-9}
			&&&&Peaks&Weibo &Baidu&Weibo &Baidu\\
			\hline
			\textcircled{\scriptsize 1}&{Sinking of a Cruise Ship} & Disaster & 06/01-06/30 (2015) & Multiple & 308.45 & 1560.4  & 320.59 & 401.48 \\
			\hline
			\textcircled{\scriptsize2}&{Chinese Stock Market Crash} & Disaster & 06/16-07/13 (2015) & Multiple & 701.71 & 420.40 & 578.77 &~74.14\\
			\hline
			\textcircled{\scriptsize3}&{AlphaGo} & High-Tech & 02/20-03/29 (2016) &Multiple & 838.12 &2337.3  & 654.89& 406.83 \\
			\hline
			\textcircled{\scriptsize4}&{Leonardo DiCaprio, Oscar Best Actor} & Entertainment & 02/20-03/09 (2016)  & Single& 2569.5& 730.82 & 1788.9  & 139.52 \\
			\hline
			\textcircled{\scriptsize5}&{Kobe Bryant's Retirement} & Sports & 04/10-04/19 (2016) & Single& 3655.3 & 2300.9 & 2274.8& 403.69\\
			\hline
			\textcircled{\scriptsize6}&{Huo and Lin Went Public with Romance$^\dag$} & Entertainment & 05/10-05/29 (2016) &Hybrid& 1535.2 & 1615.2 & 1027.1 & 289.98\\
			\hline
			\textcircled{\scriptsize7}&Brexit Referendum & Politics & 06/20-06/30 (2016) & Single& 957.16 & 2160.4 & 715.51  & 392.32 \\
			\hline
			\textcircled{\scriptsize8}& Pok\'{e}mon Go & High-Tech & 07/02-07/20 (2016) & Hybrid & 936.38 & 3652.2& 695.90 & 625.87 \\
			\hline
			\textcircled{\scriptsize9}&{The South China Sea Arbitration} & Politics & 07/10-07/19 (2016) & Single& 7671.0 & 7815.3 & 5918.2 & 1451.9\\
			\hline
		\end{tabular}
		\begin{tablenotes}\scriptsize
			
			\item[$\dag$] Actor Wallace Huo and Actress Ruby Lin went public with Romance on May 20\textsuperscript{th}, 2016.
		\end{tablenotes}
	\end{threeparttable}
\end{table*}


\subsubsection{Average Leading Time $\delta^A$ and $\delta^B$}
As visualized in Fig.~\ref{fig:heatmap}, the deviated path from the main diagonal direction indicates the lead-lag relation between the two EPTSs.
We define the \textit{Average Leading Time} $\delta^A$ of EPTSs $\mathscr E^A$ to $\mathscr E^B$ as
\begin{equation}
\delta^A = \dfrac{\sum_{m_k\in M}(i-j)}{\left|\mathscr{E}^A\right|}, \quad  i>j.
\end{equation}
where $m_k=(i,j) \in M$. The unit of $\delta^A$ corresponds to the temporal resolution of the EPTSs.

Given two media $A$ and $B$, $\delta^A$ measures the average leading time of event dissemination on medium $A$ compared with that on medium $B$, based on the alignments by $\omega$DTW-CD.

\section{Experiments}
\label{sec:verification}


\subsection{Experiment Setup}
\subsubsection{Datasets}
Our experiments are conducted on eighteen real-world event datasets from Weibo and Baidu, covering nine most popular events that occurred from 2015 to 2016. The data are extracted via internal APIs.
All the nine events covered in our datasets have provoked intensive discussions and gathered widespread attention. In addition, they are both typical events in distinct categories including disasters, high-tech stories, entertainment news, sports and politics.
The detailed information of our datasets is listed in Table~\ref{datasetsinfo}. The information includes the event name, category the events fall into, time spans, the number of peaks, the number of records for each dataset, and finally the disk size of the each dataset.

Our models are not limited to any specific language and can generalize to others. The only difference is the data preprocessing and it will be covered soon.
In this paper, we focus on Chinese, which are logograms and it is hard to process and interpret. All our event datasets are in Chinese.

\subsubsection{Data Preprocessing}
Before applying our scheme, the raw data extracted from different media are processed as follows. First of all, we filter out all irrelevant characters, such as punctuation, hyper links, etc., to format our data. Secondly, frequently used conjunctions, pronouns and prepositions are cleaned by the so called stopword-removal process. Finally, we split each record in the data set into single words, using an open-source word segmentation tool.


\subsubsection{Implementation and Parameters}
We adopt CBOW with negative sampling when implementing Word2Vec~\cite{ DBLP:conf/nips/MikolovSCCD13}. The parameters involved in TF-SW are set to be $\beta = 0.7$, with $\gamma = 0.02$ considering the nature of Chinese language, that
there are many different characters but almost no meaning changes on words.
The factor for TextRank is set to be $\theta=0.85$ by convention.
Without specification, we set each time interval to be $1$ day. The corresponding parameters for the sigmoid-like temporal weight are set as $\eta =10, \tau=2$.

\textbf{Environment:}
All the experiments are conducted on a PC device (Intel\textregistered ~ Core\texttrademark ~i7 3.5GHz, 16GB memory) and are implemented in Python 2.7.

\subsection{Verification of TF-SW}
\label{sec:very-TF-SW}

\begin{figure*}[htb]
	\begin{minipage}{.33\linewidth}
		\centering
		\includegraphics[width=\linewidth]{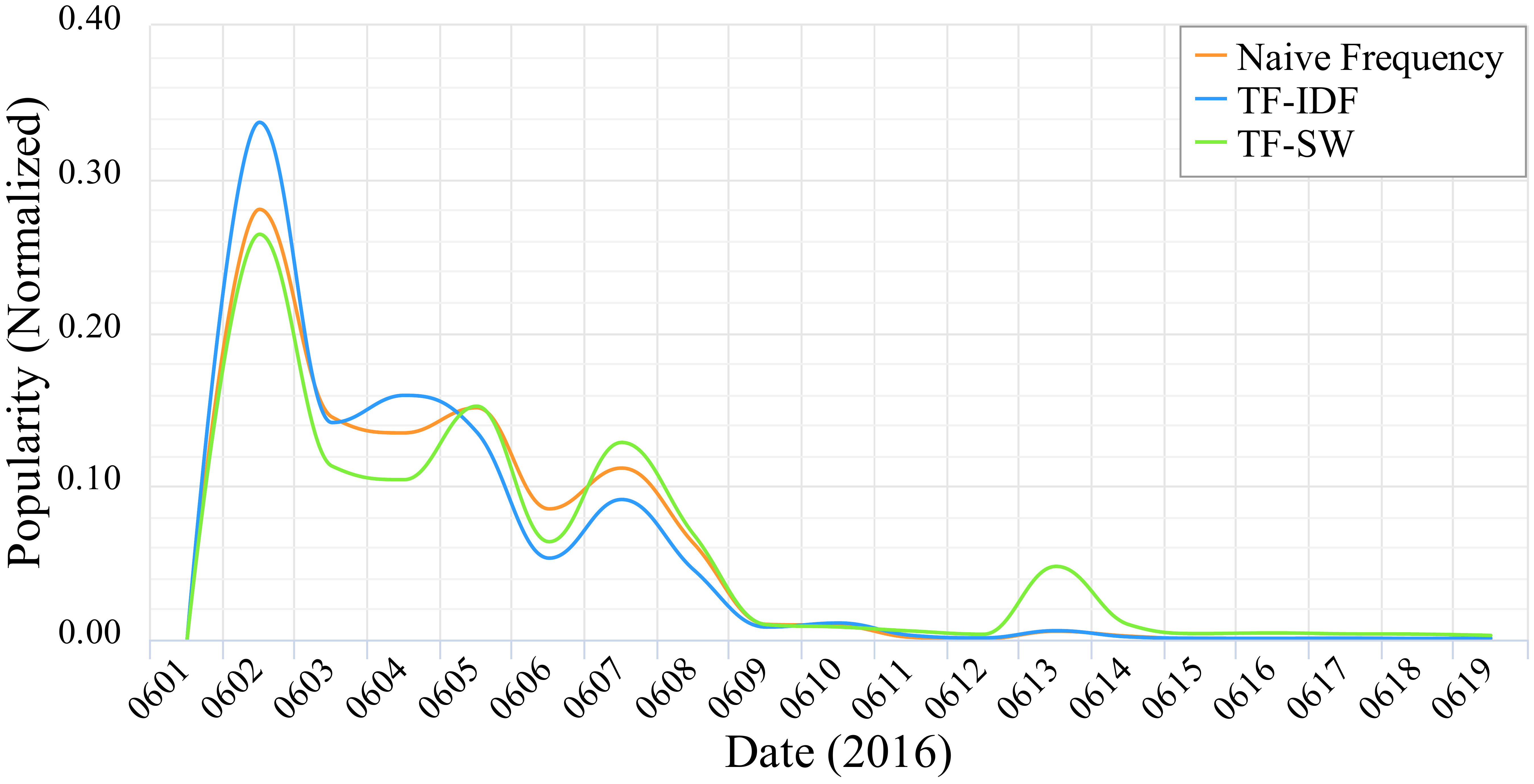}
		\caption{EPTSs of Event \textit{Sinking of a Cruise Ship} on Weibo \footnotesize{($th=10$)}}\label{fig_dfzxweibo}
	\end{minipage}%
	\hfill
	\begin{minipage}{.33\linewidth}
		\centering
		\includegraphics[width = \linewidth]{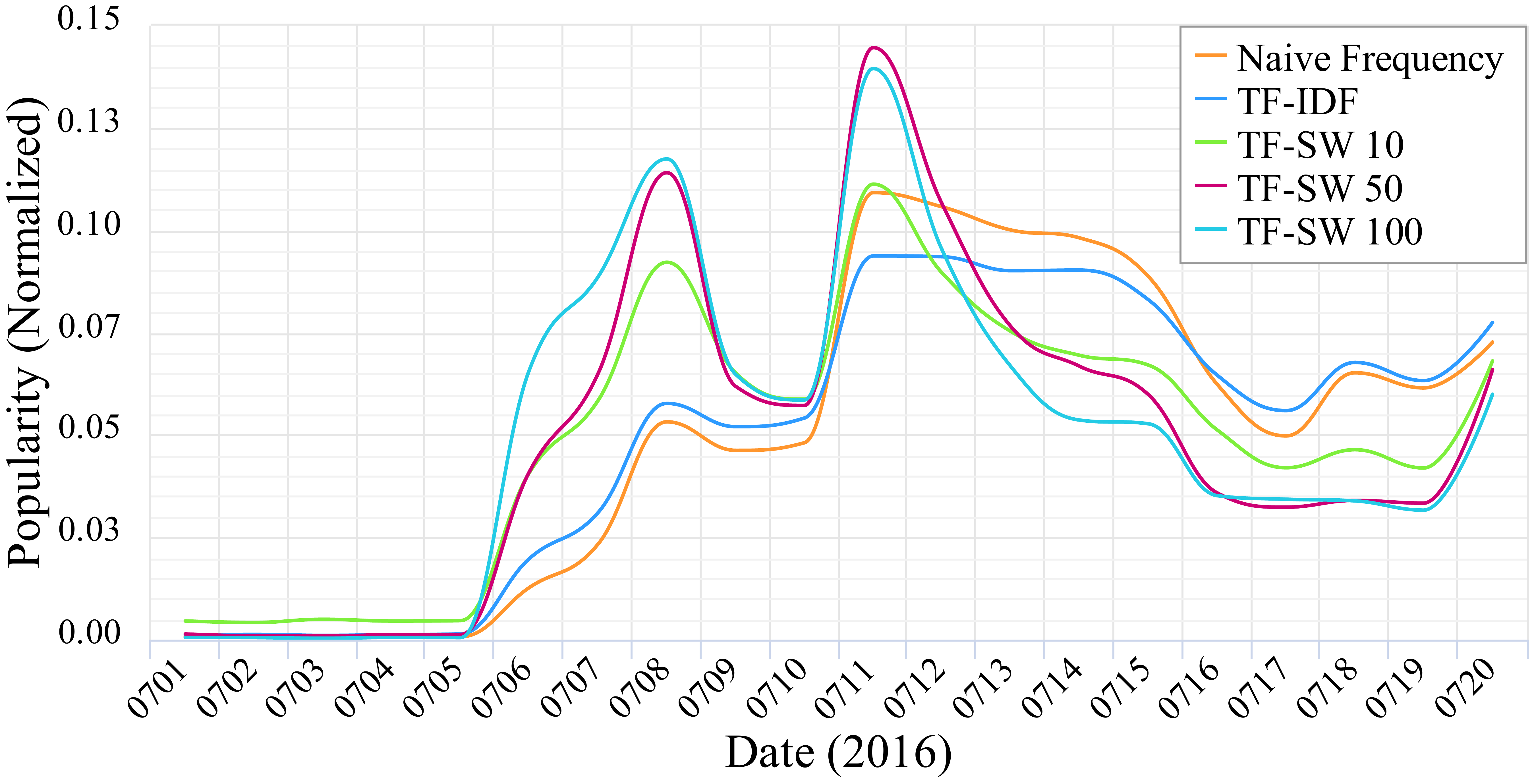}
		\caption{EPTSs of Event \textit{Pok\'{e}mon Go} on Baidu \footnotesize{(TF-SW $N$ means $th=N$.)}}
		\label{fig:pofreth}
	\end{minipage}%
	\hfill
	\begin{minipage}{.33\linewidth}
		\centering
		\includegraphics[width=\linewidth]{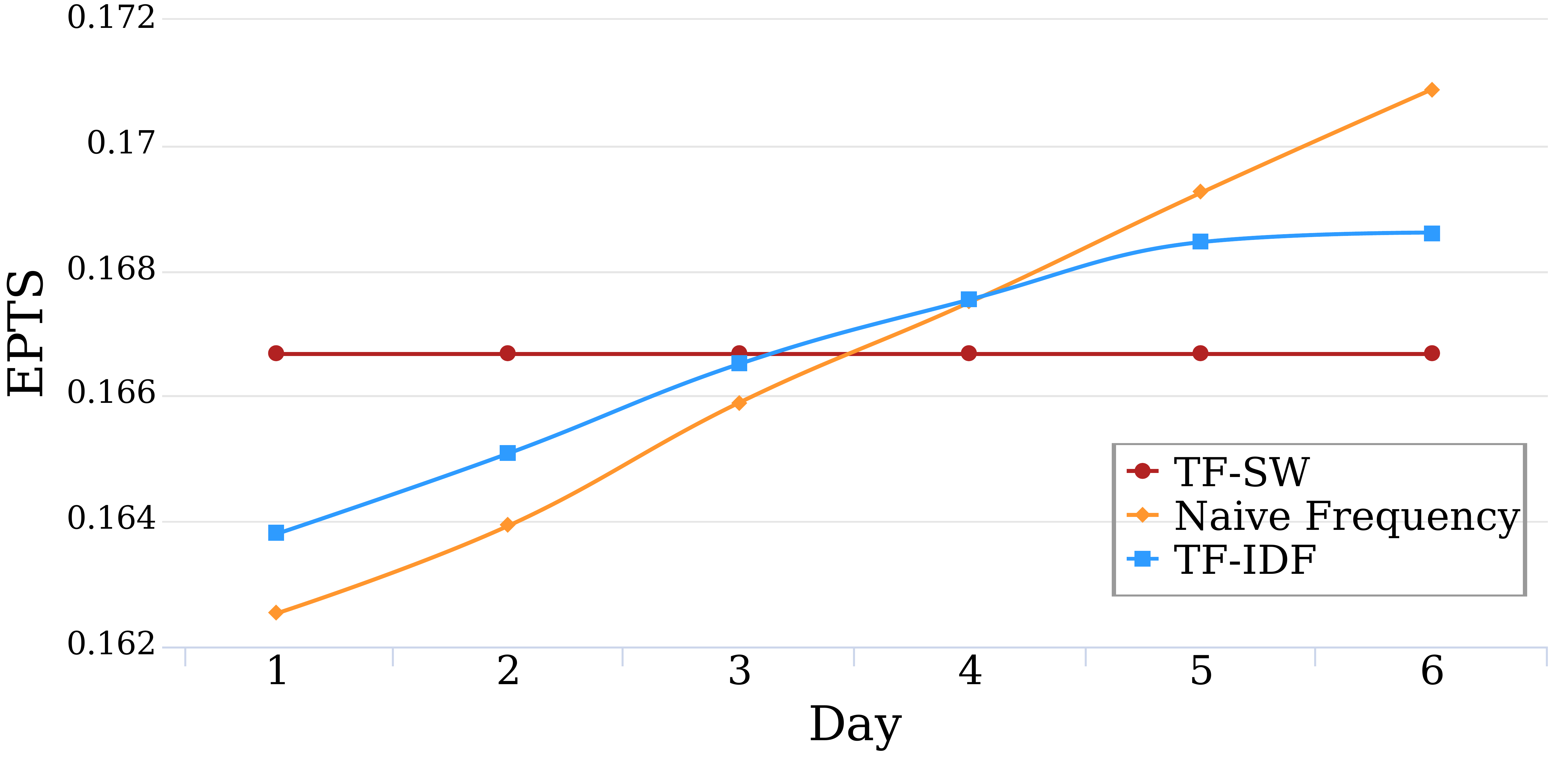}
		\caption{EPTSs on the simulated corpus}\label{fig:pocomp}
	\end{minipage}%

\end{figure*}

To evaluate the effectiveness of TF-SW, we compare the EPTS generated by our model with the EPTSs by other two baselines.

\textbf{Naive Frequency:} The popularity of a word is generated simply by word count frequency within the time interval $t_i$,
\[pop^N(E_i) = \sum\limits_{w^i_k \in E_i}fre(w^i_k).\]

\textbf{TF-IDF:} TF-IDF~\cite{WISE2016/ESAPCroPlatTreAnaSNSE} is a well developed method to evaluate relative importance of a word in a corpus and is always used as a baseline. 
The event popularity within $t_i$ is now generated by
\[pop^I(E_i)=\sum\limits_{w^i_k \in E_i}\!fre(w^i_k)\cdot\left(1+\log\frac{\left|\{R^i_j\}\right|}{\left|\{R^i_j\mid w^i_k\in R^i_j\}\right|}\right),\]
where $R_j^i$ refers to the $j$th records with timestamp $t_i$.

All the EPTSs generated by Naive Frequency and TF-IDF are normalized in the same way as TF-SW through Eqn.~\eqref{eqn:eventnormal}.

Based on the three generated EPTSs, we present a thorough discussion and comparison to validate our TF-SW model.

\subsubsection{Accuracy}
We pick up the peaks in EPTSs and backtrack what exactly happened in reality. An event is always pushed forward by series of ``little'' events and we call them sub-events, which are reflected as peaks in EPTS figures.

In the Event \textit{Capsizing of a Cruise Ship}, the real-world event evolution involves four key sub-events. On the night of June 1, 2015, the cruise ship sank in a severe thunderstorm. Such a shocking disaster raised tremendous public attention on June 2. On June 5, the ship was hoisted and set upright. A mourning ceremony was held on June 7, and on June 13, total 442 deaths and only 12 survivors were officially confirmed, which marked the end of the rescue work.


The EPTS generated by TF-SW shows four peaks, which is illustrated in Fig.~\ref{fig_dfzxweibo}.
All these peaks are highly consistent with the four key sub-events in real world, while the end of rescue work on June 13 is missed by approaches based on Naive Frequency and TF-IDF. In conclusion,  TF-SW model shows the ability to track the development of events precisely.

\begin{figure*}[htb]
	\begin{subfigure}[t]{0.23\linewidth}
		\centering
		\includegraphics[width=\linewidth]{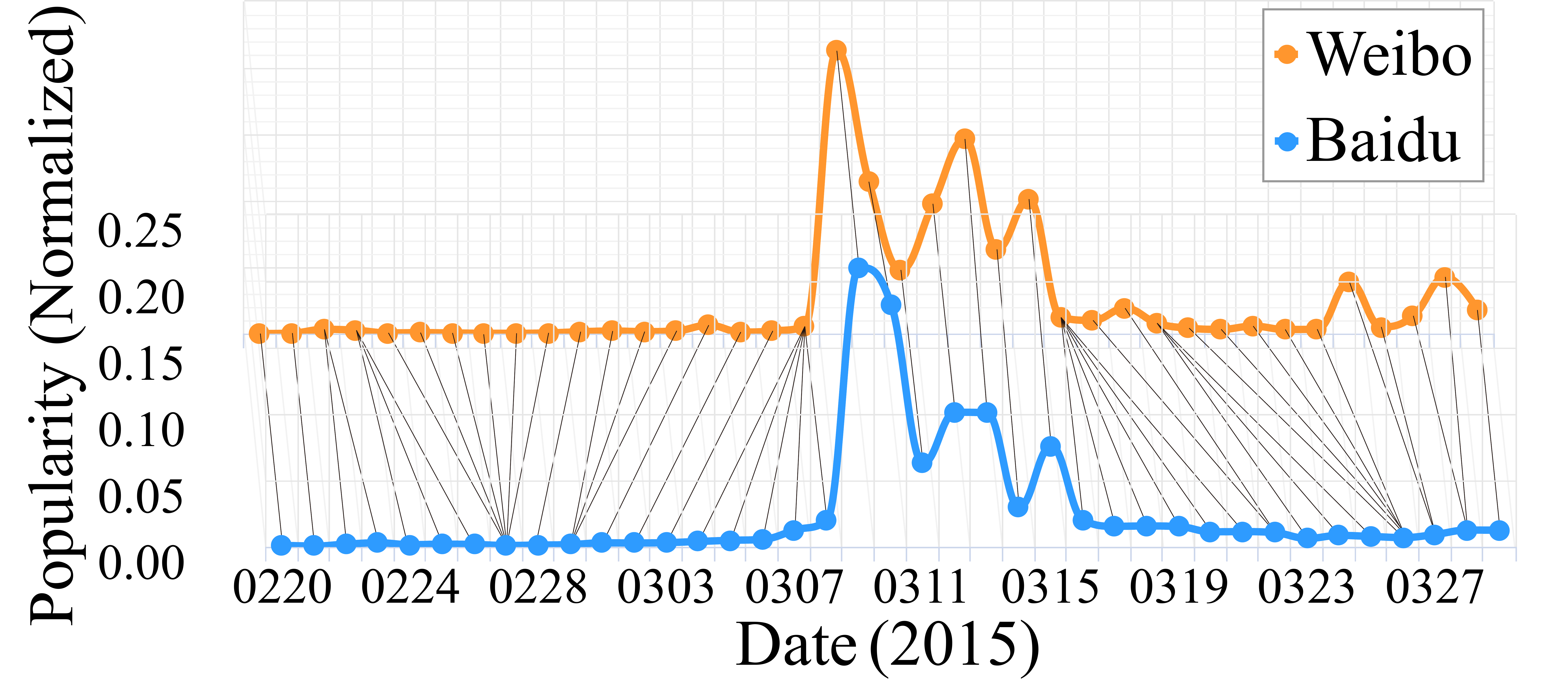}
		\caption{classic DTW}
		\label{fig:singuDTW}
	\end{subfigure}
	\hspace{1.89mm}
	\begin{subfigure}[t]{0.23\linewidth}
		\centering
		\includegraphics[width=\linewidth]{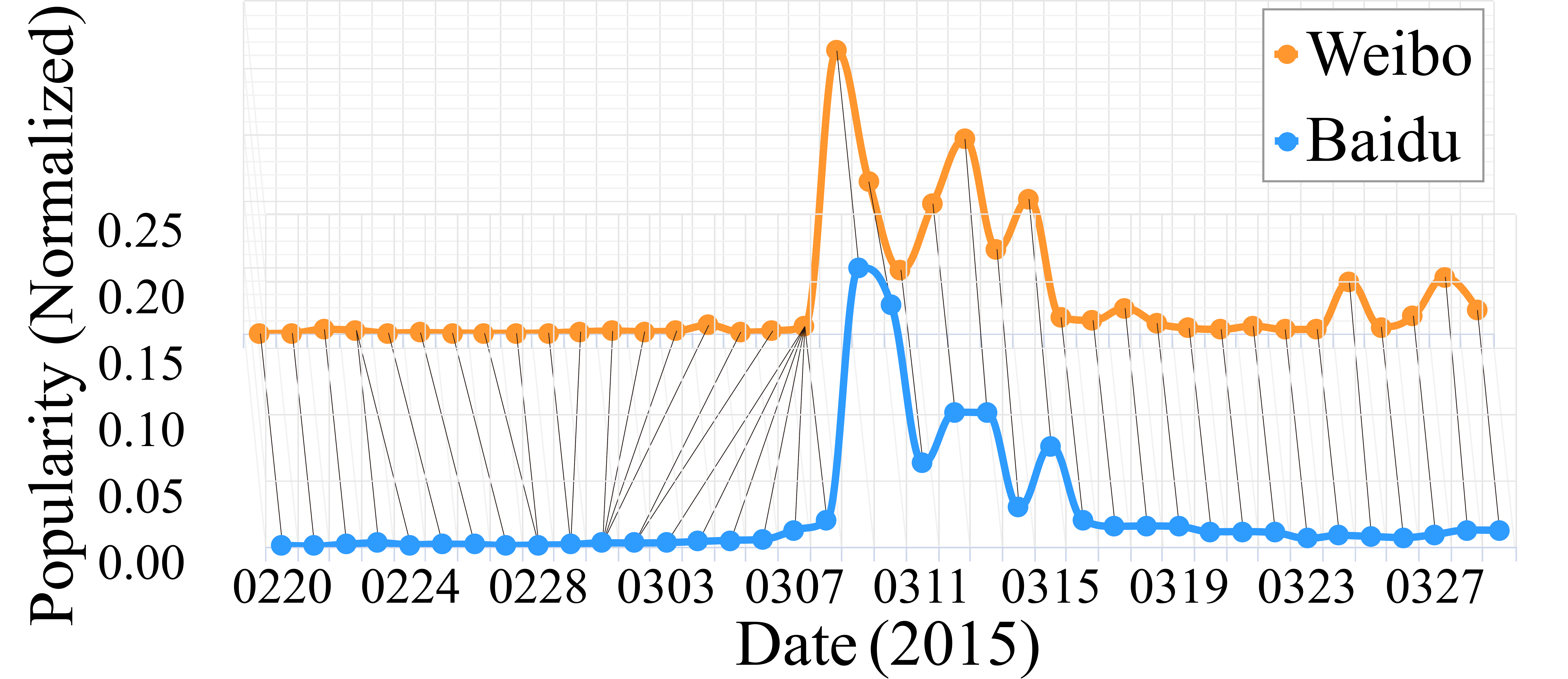}
		\caption{DTW\textsubscript{bias}  $(1.2,1,1.2)$}
		\label{fig:singuDTWbias}
	\end{subfigure}
	\hspace{1.89mm}
	\begin{subfigure}[t]{0.23\linewidth}
		\centering
		\includegraphics[width=\linewidth]{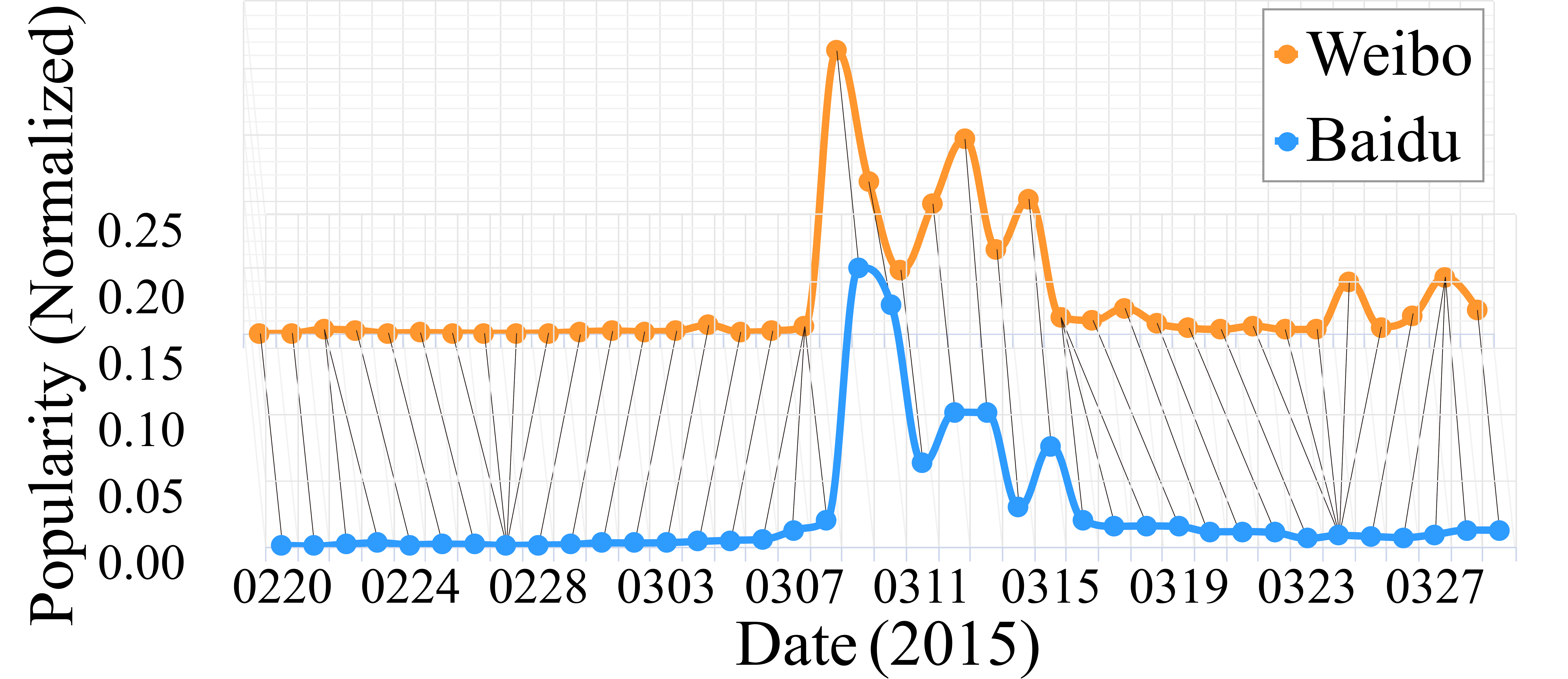}
		\caption{$\omega$DTW {\footnotesize $(\eta = 10, m =2)$}}
		\label{fig:singuwDTW}
	\end{subfigure}
	\hspace{1.89mm}
	\begin{subfigure}[t]{0.23\linewidth}
		\centering
		\includegraphics[width=\linewidth]{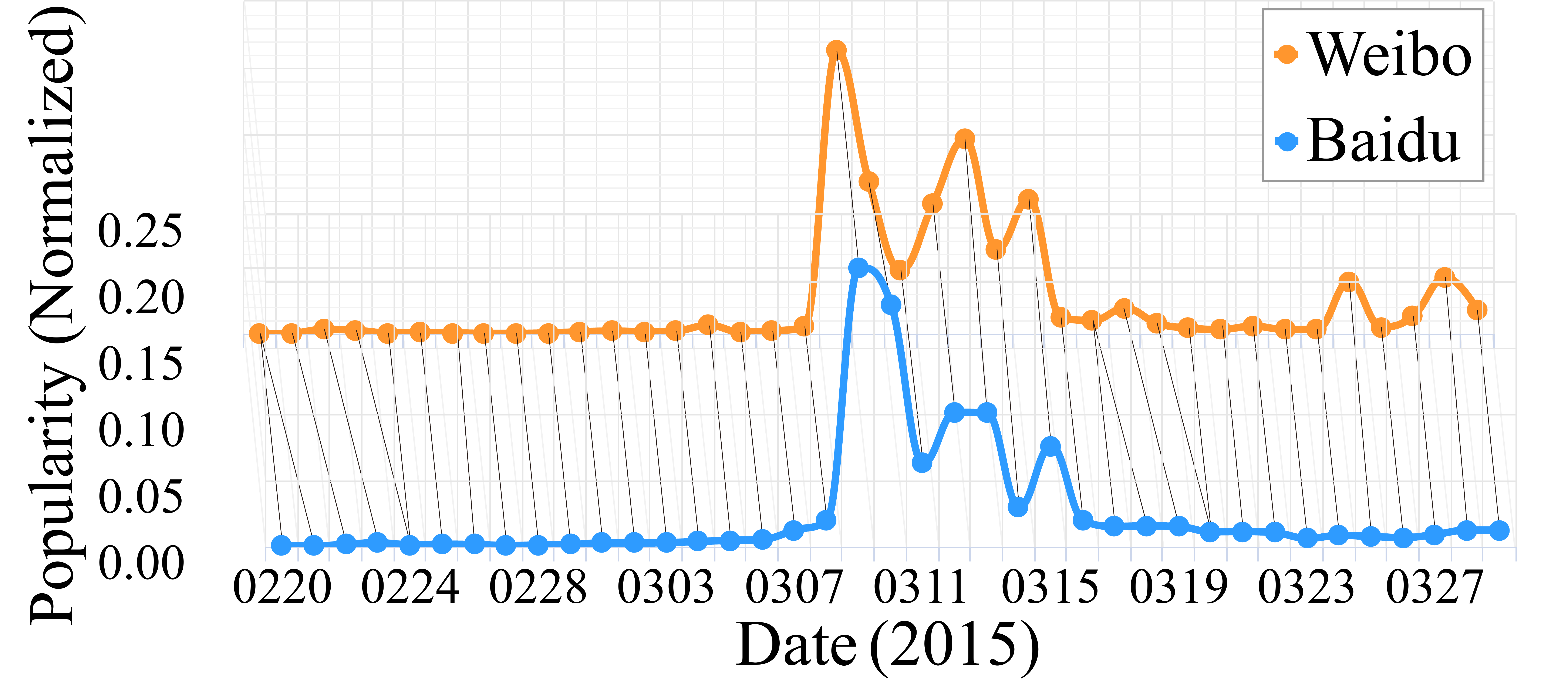}
		\caption{$\omega$DTW-CD {\footnotesize $(\eta = 10, m = 2)$}}
		\label{fig:singuwDTW-CD}
	\end{subfigure}
	\caption{Alignment results of $4$ methods on Event \textit{AlphaGo}. In this case, one datapoint is categorized as a singular point if it is matched to more than $4$ data points from the other EPTS.}
	\label{fig_singuexp}
\end{figure*}

\subsubsection{Sensitivity to Burst Phases}
In TF-SW, a frequency threshold $th$ is introduced to filter the noisy words due to the observed \textit{Long Tail} in Fig.~\ref{fig:wordcntfre}.
To demonstrate the function of our model on highlighting contributive words, we conduct experiments on TF-SW with distinct $th$ on Event \textit{Pok\'{e}mon Go} and compare the results with the other two baselines.
%
%


Compared with the baselines, our model are more sensitive to the burst phases of an event, as is shown in Fig.~\ref{fig:pofreth}, especially on data points 07/06, 07/08, and 07/11. The event popularity on these days are larger than those obtained by Naive Frequency and TF-IDF. In another word, the EPTSs generated through TF-SW rises faster, more significant in peaks, and are more sensitive to breaking news which enables the model to capture the burst phases more precisely.

From three EPTSs of TF-SW with different $th$, it is shown that TF-SW is more sensitive to the burst of events with a higher $th$ value, as is shown by the data point 07/06.

An event whose EPTS rises fast at some data points possesses the potential to draw wider attention. It is reasonable for a popularity model not only to depict the current state of event popularity, but also take the potential future trends into consideration.
In this way, a quick response to the burst phases of an event is more valuable for real-world applications. This advantage of our model can lead to a powerful technique for first story detection on ongoing events.

\subsubsection{Superior Robustness to Noise}
To verify whether our model can effectively filter out noisy words, we further implement an experiment on a simulated corpus.

We first extract 50K Baidu queries with the highest frequency in the corpus of Event \textit{Kobe's Retirement} and make them as the base data for a 6-day simulated corpus. Then we randomly pick noisy queries from Internet that are not relevant to Event \textit{Kobe's Retirement} at all. The amount of noisy queries is listed in Table~\ref{tab:noiserec}.

\begin{table}[htbp]
	\caption{Number of noisy records added to each day}
	\label{tab:noiserec}
	\centering
	\begin{tabular}{ c |cccccc}
		\hline
		Day & 1 & 2 & 3 & 4 & 5 & 6  \\
		\hline
		\# (k) & 0.000 & 1.063 & 2.235 & 3.507 & 4.689 & 6.026 \\
		\hline
	\end{tabular}
\end{table}

Since each day's base data are identical, say that 50K queries, a good model is supposed to filter noisy queries out and generate an EPTS with all identical data points, which form a horizontal line in X-Y plane. EPTSs generated by TF-SW, Naive Frequency and TF-IDF are shown in Fig.~\ref{fig:pocomp}. It is shown that TF-SW successfully filters out the noise and generates the EPTS which is a horizontal line and captures the real event popularity, while the other two methods Naive Frequency and TF-IDF are obviously effected by the noisy queries and generate EPTSs that cannot accurately reflect the event popularity.


\subsection{Verification of $\omega$DTW-CD}

\begin{figure*}[ht]
	\begin{subfigure}[t]{0.23\linewidth}
		\centering
		\includegraphics[width=\linewidth]{dfzx-ddtw-3d.pdf}
		\caption{DDTW}
		\label{fig:farmDTW}
	\end{subfigure}
	\hfill
	\begin{subfigure}[t]{0.23\linewidth}
		\centering
		\includegraphics[width=\linewidth]{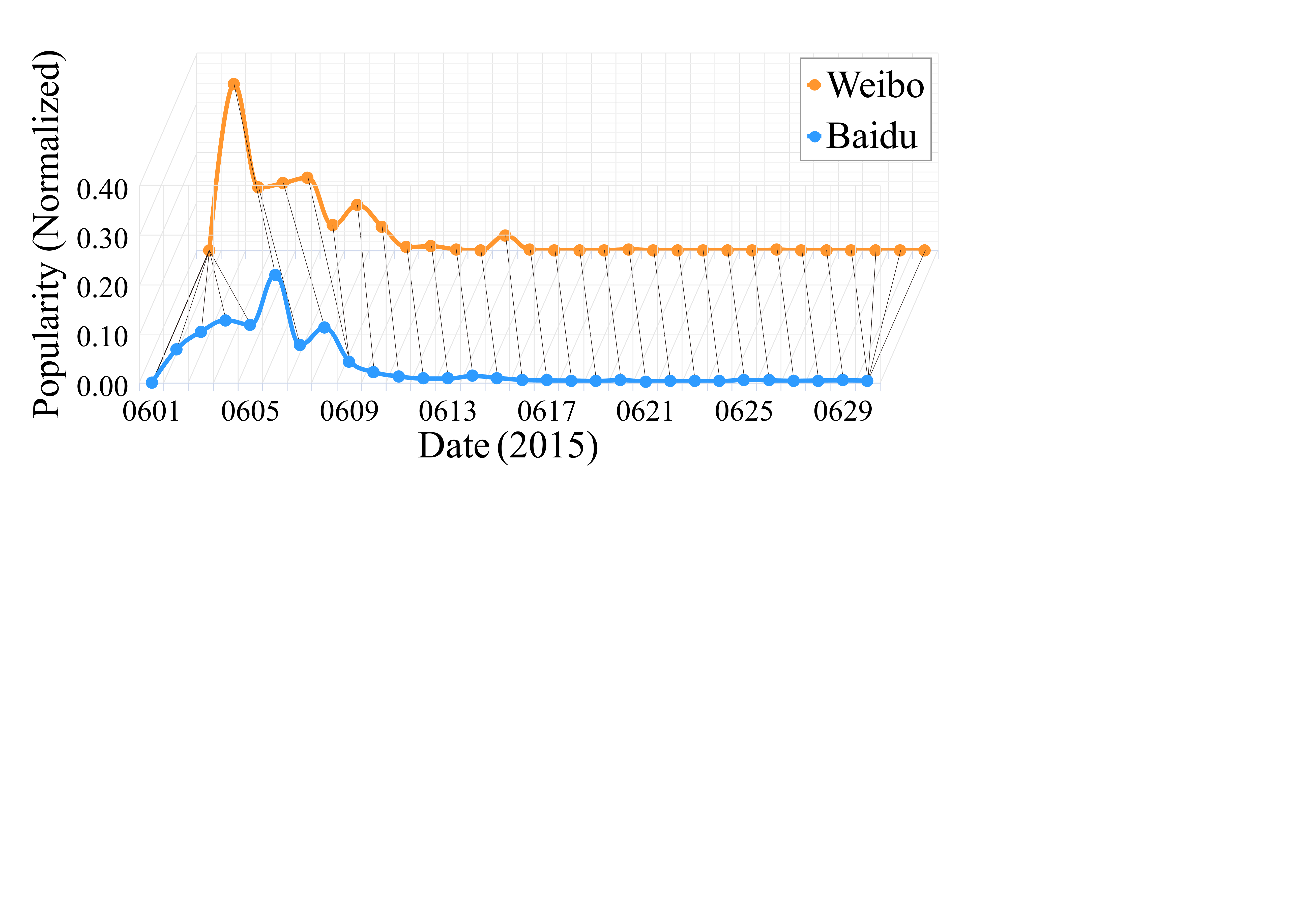}
		\caption{DDTW\textsubscript{bias} $(1.2,1,1.2)$}
		\label{fig:farmDTWbias}
	\end{subfigure}
	\hfill
	\begin{subfigure}[t]{0.23\linewidth}
		\centering
		\includegraphics[width=\linewidth]{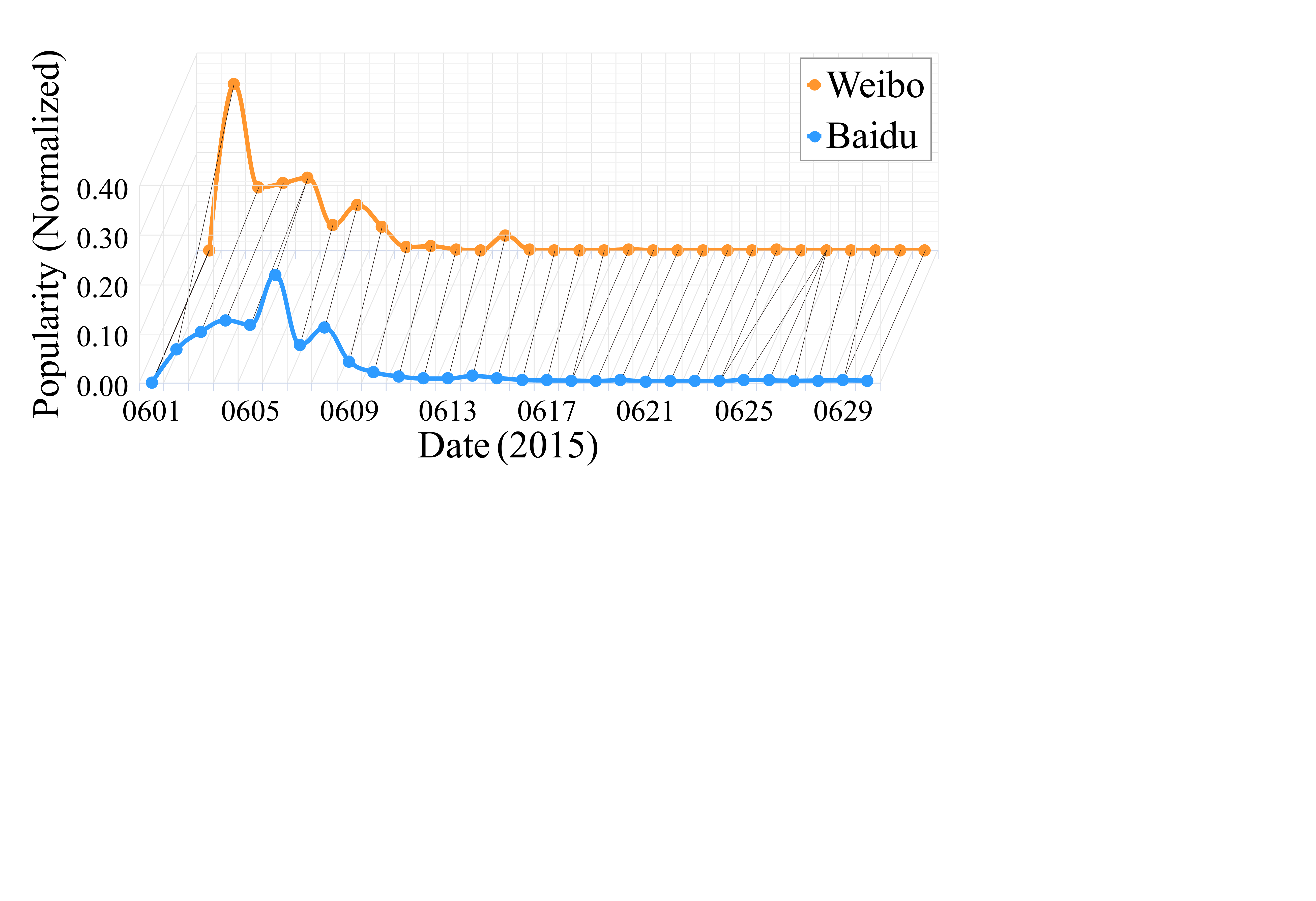}
		\caption{$\omega$DDTW {\footnotesize $(\eta = 5, m =3.2)$}}
		\label{fig:farmwDTW}
	\end{subfigure}
	\hfill
	\begin{subfigure}[t]{0.23\linewidth}
		\centering
		\includegraphics[width=\linewidth]{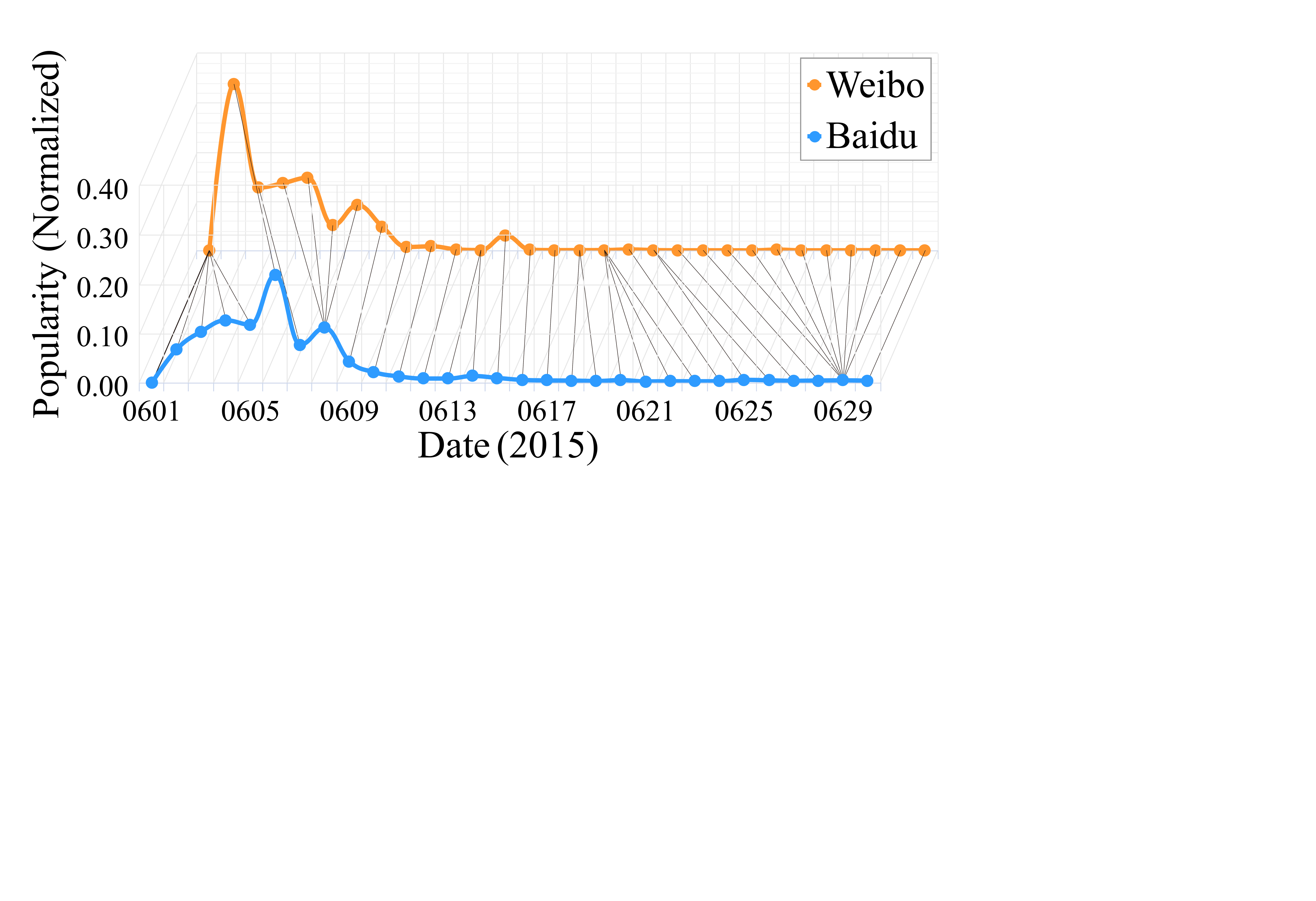}
		\caption{DTW-CD}
		\label{fig:farmDTW-CD}
	\end{subfigure}
	\caption{Alignment results of four methods on Event \textit{The Capsizing of a Cruise Ship}}
	\label{fig_farmexp}
\end{figure*}

To demonstrate the effectiveness of $\omega$DTW-CD, we compare it with seven different DTW extensions listed below.

\begin{itemize}
\item \textit{DTW} is the DTW method with Euclidean distance.
\item \textit{DDTW}~\cite{keogh2001derivative} is the Derivative DTW which replaces the Euclidean distance with the difference of estimated derivatives of the data points in EPTSs.
\item \textit{DTW\textsubscript{bias} \& DDTW\textsubscript{bias}} are the extended DTW and DDTW respectively with a bias towards the diagonal direction. The cumulated cost matrices are obtained by
\[d^*_{i,j}=d_{i,j}+\min\{ b_1 \cdot d^*_{i-1,j}, b_2 \cdot d^*_{i-1,j-1}, b_3 \cdot d_{i,j-1}\},\]
instead of Eqn.~\eqref{eq:accugen}, where $(b_1,b_2,b_3) $ is a vector of positive real numbers proposed by~\cite{sakoe1978dynamic}. 
\item \textit{$\omega$DTW \& $\omega$DDTW} are the temporally weighted DTW and DDTW, where the sigmoid-like temporal weight defined by Eqn.~\eqref{eq:temweight} is introduced to the cost matrices.
\item \textit{DTW-CD} is a simplification of \textit{wDTW-CD} that implements only $dist^C$ 
without temporal weight $\omega$.
\end{itemize}

\begin{figure*}
	\centering
	\includegraphics[width=\textwidth]{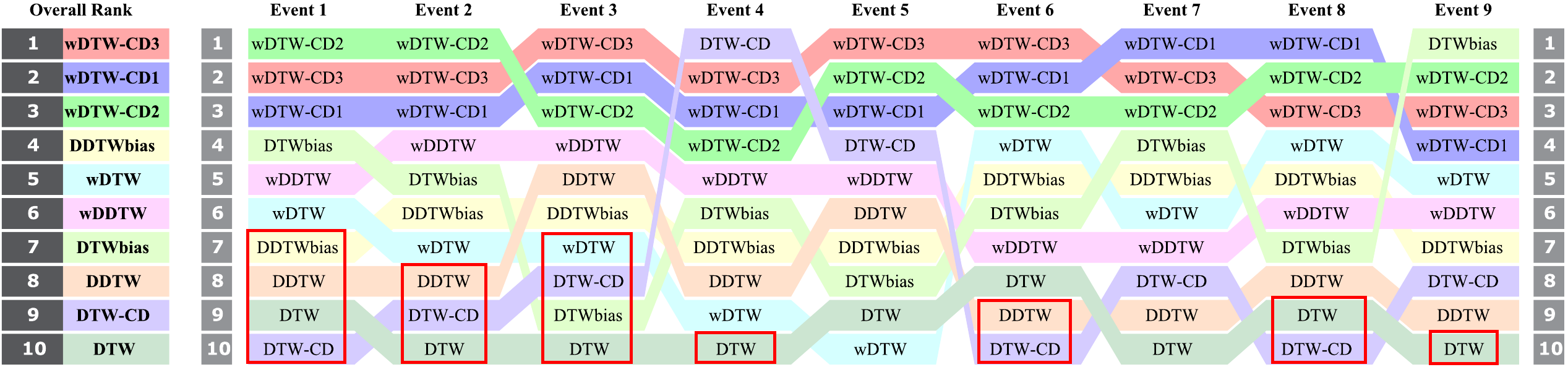}
	\caption{Ranking visualization
		of grades for 10
		methods on nine real-world events.
	}
	\label{fig:rankingchart}
\end{figure*}

\subsubsection{Singularity}
Fig.~\ref{fig_singuexp} visualizes the results generated by DTW, DTW\textsubscript{bias}, $\omega$DTW and our proposed model.
	Classic DTW and DTW\textsubscript{bias} severely suffer the problem of singularity.
	Compared with $\omega$DTW, $\omega$DTW-CD presents better and more stable performance when aligning the time series with sharp fluctuations.
	In general, our model is capable of avoiding the singularity problem by involving the derivative differences.

	\subsubsection{Far-Match}
	Considering the fact that the time difference between two aligned sub-event can barely exceed two days, far-match exists in the alignment generated by DDTW, DDTW\textsubscript{bias} and DTW-CD in Fig.~\ref{fig_farmexp}, but not in our results in Fig.~\ref{fig:3dalign}.
	Thus, the sigmoid-like temporal weight introduced to our model helps effectively avoid the far-match problem.


	\subsubsection{Overall Performance}
	All the comparison results on the eighteen real-world datasets are illustrated in Fig.~\ref{fig:rankingchart}, where each color corresponds to a method, each method are ranked respectively for each event, and methods with higher grades are ranked on the top. Results facing \textit{singularity} or \textit{far-match} are marked by red boxes.
	The performances are graded under the following criteria.

	\begin{itemize}
		\item The grades are given to show the relative performances among different methods only regarding one event.
		\item The method that does not suffer from \textit{singularity} or \textit{far-match} has higher grades than the one that does.
		\item  The methods giving same alignment results are further graded considering their complexity.
	\end{itemize}

	In comparison with existing variants of DTW as well as the reduced version of our method, $\omega$DTW-CD achieves improvements on both performance and robustness on alignment generation and successfully conquers the problem of \emph{singularity} and \emph{far match}.
	Results shows that the event phase distance, estimated derivative difference, and the sigmoid-like temporal weight simultaneously contribute to the performance enhancement of $\omega$DTW-CD.

	Moreover, with parameter $\eta$ and $\tau$, our model is flexible to different temporal resolutions and to events of distinct popularity features. In Fig.~\ref{fig:rankingchart}, $\omega$DTW-CD$_1$ corresponds to $\eta=5,\ \tau =3.2$. $\eta=10,\ \tau =2$ is for $\omega$DTW-CD$_2$. $\eta=5,\ \tau =2.2$ is for $\omega$DTW-CD$_3$. The results show the strong ability of $\omega$DTW-CD to handle specific events.

	\subsection{Case Studies}
	\label{sec:casestudy}
	We apply \textsc{DancingLines} on the real-world events to illustrate the application potentials of the scheme.





	\subsubsection{Case 1: Superiority from Flexible Temporal Resolution}
\begin{figure*}[ht]
		\begin{minipage}{.33\linewidth}
			\centering
			\includegraphics[width=\linewidth]{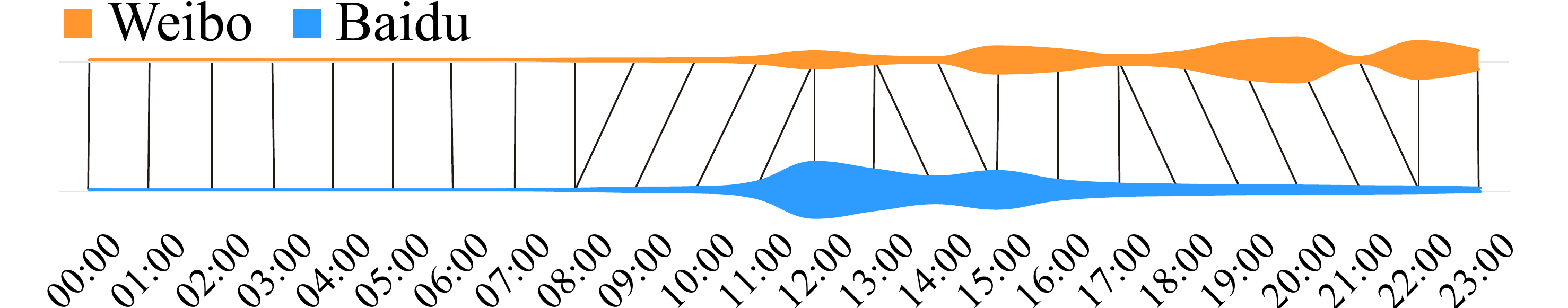}
			\caption{Aligned EPTSs {\footnotesize Event \textit{AlphaGo} on March 9}}\label{alpha-hour}
		\end{minipage}%
		\hfill
		\begin{minipage}{.33\linewidth}
			\centering
			\includegraphics[width = \linewidth]{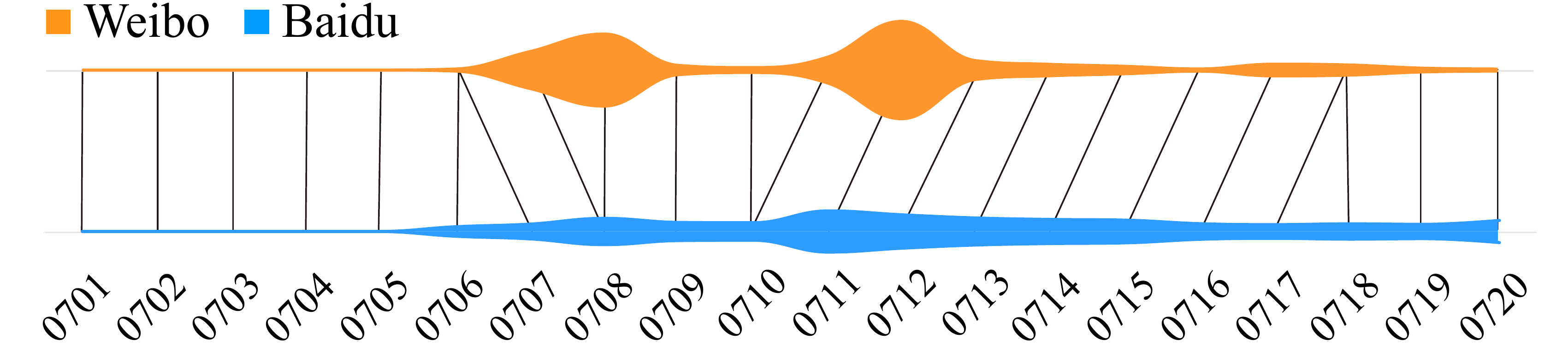}
			\caption{Aligned EPTS {\footnotesize Event \textit{Pok\`{e}mon Go}}}
			\label{pokemongo-align}
		\end{minipage}%
		\hfill
		\begin{minipage}{.33\linewidth}
			\centering
			\includegraphics[width=\linewidth]{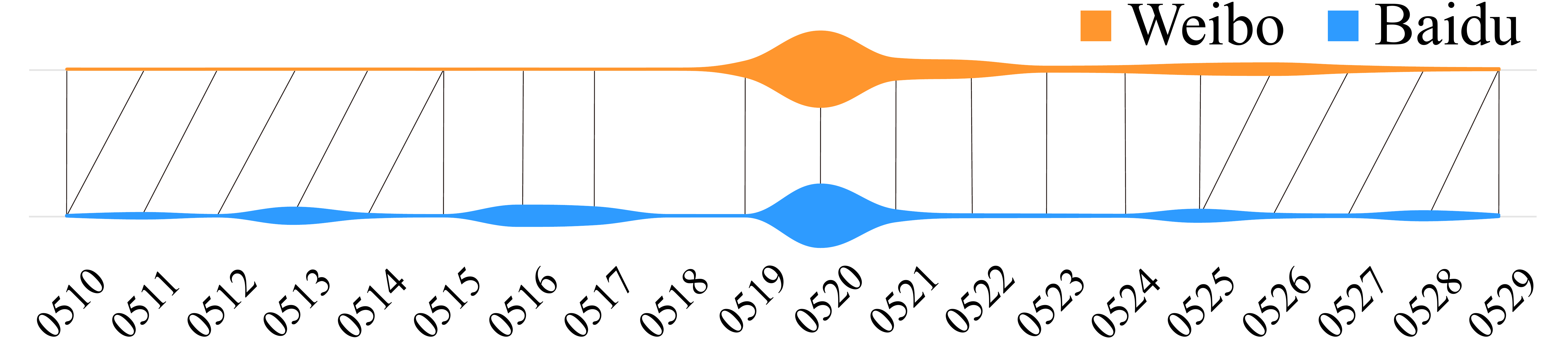}
			\caption{Lead-lag Stripes ({\footnotesize  Event \textit{Huo and Lin Went Public with Romance}})}
			\label{fig:linxinruali}
		\end{minipage}%
	\end{figure*}

	The EPTSs generated at different temporal resolutions contain knowledge from different granularity regarding event popularity.
	We conduct experiments with the length of each time interval set to be an hour.
	An example on Event \textit{AlphaGo} is shown in Fig.~\ref{alpha-hour}.
	The EPTS at the temporal resolution of one day can capture and reveal the game-level information of Event \textit{AlphaGo}, while if we choose smaller time resolution, we can obtain more details about dissemination.
	According to the alignment shown in Fig.~\ref{alpha-hour}, at the very beginning, event \textit{AlphaGo} arouse noticeable public attention on Baidu.
	After several hours, the event popularity on Weibo started increasing, while that on Baidu decayed.
	This flexibility to choose temporal resolutions satisfies various users' applications.




	\subsubsection{Case 2: Focusing on Events and Metrics}
The three metrics based on aligned cross-platform event phases shed lights on related analysis. Taking Event \textit{AlphaGo} and Event \textit{Pok\`{e}mon Go} as an example, \textit{Time-Irrelevant Altitude Similarity} of the two events are $\psi_A = 0.829$ and $\psi_A = 0.660$. These values are in accordance with Fig.~\ref{fig:singuwDTW-CD} and Fig.~\ref{pokemongo-align}, that Weibo and Baidu were coupled more closely in disseminating Event \textit{AlphGo}.


Considering Event \textit{Chinese Stock Market Crash}, the \textit{Time-Irrelevant Shape Similarity} $\psi_S$ is the minimum among the nine events. This agrees with the fact that shareholders are willing to express their feelings instantly on social netwok, Weibo. 

Events in the category of disasters (\emph{Sinking of a Cruise Ship} and \emph{Chinese Stock Market Crash}) rank the top on \textit{Average Leading Time} of Weibo $\delta^W$, which indicates that these events gain popularity more quickly on social networks like Weibo.



	Comparison on different metric values of specific events between media can lead to insight on event properties. Findings on properties of media can be also acquired by comparing categories and properties of events that generate similar results.

	\subsubsection{Case 3: On Multiple Information Sources}
	In common sense, social networks like Weibo should react more quickly than search engines on events of entertainment.
	We then look deeper into the two aligned EPTSs. The EPTS for Baidu, which takes the lead according to our metrics, contains periodic peaks every 3--5 days, as shown in Fig.~\ref{fig:linxinruali}. This pattern may indicate deliberate paparazzi hypes.
	The only breakout on Weibo is aligned to one of peaks in Baidu EPTS. The peak fits well to the periodic pattern. Therefore, a possible explanation is that Event \emph{Huo and Lin Went Public with Romance} is an event deliberately promoted and successfully hyped.

	\section{Conclusion}\label{sec:conclusion}
	
	In this paper, we have studied how to quantify and interpret event popularity between pairwise text media with an innovative scheme named \textsc{DancingLines}. To address the popularity quantification issue, we utilize TextRank and Word2Vec to transform the corpus into a graph and project the words into vectors, which are covered in TF-SW model. To furthermore interpret the temporal warp between two EPTSs, we propose $\omega$DTW-CD which includes a novel compound distance measurement and temporal weight, to generate alignments of EPTSs. Besides that, three metrics are proposed to bridge deeper analysis. Experimental results on eighteen real-world event datasets from Weibo and Baidu validate the effectiveness and applicability of our scheme.


%
%

	\bibliographystyle{IEEEtran}

	\bibliography{IEEEabrv,sigproc}

\begin{thebibliography}{10}
\providecommand{\url}[1]{#1}
\csname url@samestyle\endcsname
\providecommand{\newblock}{\relax}
\providecommand{\bibinfo}[2]{#2}
\providecommand{\BIBentrySTDinterwordspacing}{\spaceskip=0pt\relax}
\providecommand{\BIBentryALTinterwordstretchfactor}{4}
\providecommand{\BIBentryALTinterwordspacing}{\spaceskip=\fontdimen2\font plus
\BIBentryALTinterwordstretchfactor\fontdimen3\font minus
  \fontdimen4\font\relax}
\providecommand{\BIBforeignlanguage}[2]{{%
\expandafter\ifx\csname l@#1\endcsname\relax
\typeout{** WARNING: IEEEtran.bst: No hyphenation pattern has been}%
\typeout{** loaded for the language `#1'. Using the pattern for}%
\typeout{** the default language instead.}%
\else
\language=\csname l@#1\endcsname
\fi
#2}}
\providecommand{\BIBdecl}{\relax}
\BIBdecl

\bibitem{DBLP:conf/icde/LiLKC12}
R.~Li, K.~H. Lei, R.~Khadiwala, and K.~Chang, ``Tedas: A twitter-based event
  detection and analysis system,'' in \emph{ICDE}, 2012, pp. 1273--1276.

\bibitem{DBLP:conf/kdd/LinWHY13}
S.~Lin, F.~Wang, Q.~Hu, and P.~Yu, ``Extracting social events for learning
  better information diffusion models,'' in \emph{{KDD}}, 2013, pp. 365--373.

\bibitem{DBLP:conf/icdm/WangTYLMCHH15}
J.~Wang, W.~Tong, H.~Yu, M.~Li, X.~Ma, H.~Cai, T.~Hanratty, and J.~Han,
  ``Mining multi-aspect reflection of news events in twitter: Discovery,
  linking and presentation,'' in \emph{{ICDM}}, 2015, pp. 429--438.

\bibitem{DBLP:journals/vldb/Zhou014}
X.~Zhou and L.~Chen, ``Event detection over twitter social media streams,''
  \emph{VLDB J.}, vol.~23, no.~3, pp. 381--400, 2014.

\bibitem{ASNets}
X.~Cao and Y.~Yu, ``Asnets: {A} benchmark dataset of aligned social networks
  for cross-platform user modeling,'' in \emph{{CIKM}}, 2016, pp. 1881--1884.

\bibitem{DBLP:conf/nips/MikolovSCCD13}
T.~Mikolov, I.~Sutskever, K.~Chen, G.~S. Corrado, and J.~Dean, ``Distributed
  representations of words and phrases and their compositionality,'' in
  \emph{{NIPS}}, 2013, pp. 3111--3119.

\bibitem{mihalcea2004textrank}
R.~Mihalcea and P.~Tarau, ``Textrank: Bringing order into texts,'' in
  \emph{{EMNLP}}, 2004, pp. 404--411.

\bibitem{wikitopics}
B.~Ahn, B.~Van~Durme, and C.~Callison-Burch, ``Wikitopics: What is popular on
  wikipedia and why,'' in \emph{Proceedings of the Workshop on Automatic
  Summarization for Different Genres, Media, and Languages}, 2011, pp. 33--40.

\bibitem{cikmevent1}
Y.~Rong, Q.~Zhu, and H.~Cheng, ``A model-free approach to infer the diffusion
  network from event cascade,'' in \emph{{CIKM}}, 2016, pp. 1653--1662.

\bibitem{DBLP:conf/kdd/LeeLM13}
P.~Lee, L.~V.~S. Lakshmanan, and E.~E. Milios, ``Keysee: supporting keyword
  search on evolving events in social streams,'' in \emph{{KDD}}, 2013, pp.
  1478--1481.

\bibitem{WISE2016/ESAPCroPlatTreAnaSNSE}
Y.~Tang, P.~Ma, B.~Kong, W.~Ji, X.~Gao, and X.~Peng, ``{ESAP}: A novel approach
  for cross-platform event dissemination trend analysis between social network
  and search engine,'' in \emph{{WISE}}, 2016, pp. 489--504.

\bibitem{liu2016breaking}
N.~Liu, H.~An, X.~Gao, H.~Li, and X.~Hao, ``Breaking news dissemination in the
  media via propagation behavior based on complex network theory,''
  \emph{PHYSICA A}, vol. 453, pp. 44--54, 2016.

\bibitem{DBLP:journals/amc/AbilhoaC14}
W.~D. Abilhoa and L.~N. de~Castro, ``A keyword extraction method from twitter
  messages represented as graphs,'' \emph{Applied Mathematics and Computation},
  vol. 240, pp. 308--325, 2014.

\bibitem{jin2016labelling}
Z.~Jin, Q.~Li, C.~Wang, D.~D. Zeng, and L.~Wang, ``Labelling topics in weibo
  using word embedding and graph-based method,'' in \emph{{ICISE}}, 2016, pp.
  34--37.

\bibitem{DBLP:conf/acl/ZhaoJHSALL11}
W.~Zhao, J.~Jiang, J.~He, Y.~Song, P.~Achananuparp, E.~Lim, and X.~Li,
  ``Topical keyphrase extraction from twitter,'' in \emph{{ACL}}, 2011, pp.
  379--388.

\bibitem{DBLP:journals/tomccap/BaoXMH15}
B.~Bao, C.~Xu, W.~Min, and M.~S. Hossain, ``Cross-platform emerging topic
  detection and elaboration from multimedia streams,'' \emph{TOMCCAP}, vol.~11,
  no.~4, p.~54, 2015.

\bibitem{firststory}
M.~Osborne, S.~Petrovic, R.~McCreadie, C.~Macdonald, and I.~Ounis, ``Bieber no
  more: First story detection using twitter and wikipedia,'' in \emph{SIGIR
  2012 Workshop on Time-aware Information Access}, 2012.

\bibitem{DBLP:journals/tkde/ZhouLZM16}
X.~Zhou, X.~Liang, H.~Zhang, and Y.~Ma, ``Cross-platform identification of
  anonymous identical users in multiple social media networks,'' \emph{TKDE},
  vol.~28, no.~2, pp. 411--424, 2016.

\bibitem{Giummol2013}
F.~Giummol{\`e}, S.~Orlando, and G.~Tolomei, ``A study on microblog and search
  engine user behaviors: How twitter trending topics help predict google hot
  queries,'' \emph{HUMAN}, vol.~2, no.~3, p. 195, 2013.

\bibitem{Kwak2010}
H.~Kwak, C.~Lee, H.~Park, and S.~Moon, ``What is twitter, a social network or a
  news media?'' in \emph{WWW}, 2010, pp. 591--600.

\bibitem{linkpredhan}
J.~Zhang, J.~Chen, S.~Zhi, Y.~Chang, P.~S. Yu, and J.~Han, ``Link prediction
  across aligned networks with sparse and low rank matrix estimation,'' in
  \emph{ICDE}, 2017, pp. 971--982.

\bibitem{sakoe1978dynamic}
H.~Sakoe and S.~Chiba, ``Dynamic programming algorithm optimization for spoken
  word recognition,'' \emph{{IEEE} Trans. Acoust., Speech, Signal Process.},
  vol.~26, no.~1, pp. 43--49, 1978.

\bibitem{maustime}
V.~Maus, G.~C{\^a}mara, R.~Cartaxo, A.~Sanchez, F.~Ramos, and G.~Queiroz, ``A
  time-weighted dynamic time warping method for land-use and land-cover
  mapping,'' \emph{J-STARS}, vol.~9, no.~8, pp. 3729--3739, 2016.

\bibitem{cikmdtw1}
H.~A. Dau, N.~Begum, and E.~Keogh, ``Semi-supervision dramatically improves
  time series clustering under dynamic time warping,'' in \emph{{CIKM}}, 2016,
  pp. 999--1008.

\bibitem{keogh2001derivative}
E.~J. Keogh and M.~J. Pazzani, ``Derivative dynamic time warping,'' in
  \emph{{SDM}}, 2001, pp. 1--11.

\bibitem{jeong2011weighted}
Y.-S. Jeong, M.~K. Jeong, and O.~A. Omitaomu, ``Weighted dynamic time warping
  for time series classification,'' \emph{Pattern Recogn.}, vol.~44, no.~9, pp.
  2231--2240, 2011.

\bibitem{silva2016effect}
D.~F. Silva, G.~E. Batista, and E.~Keogh, ``On the effect of endpoints on
  dynamic time warping,'' in \emph{{SIGKDD Workshop on Mining Data and Learning
  from Time Series}}, 2016.

\bibitem{zipf}
M.~Newman, ``Power laws, pareto distributions and zipf's law,'' \emph{Contemp.
  Phys.}, vol.~46, no.~5, pp. 323--351, 2005.

\bibitem{DBLP:journals/tvcg/ByronW08}
L.~Byron and M.~Wattenberg, ``Stacked graphs - geometry {\&} aesthetics,''
  \emph{TVCG}, vol.~14, no.~6, pp. 1245--1252, 2008.

\bibitem{simhash02}
M.~S. Charikar, ``Similarity estimation techniques from rounding algorithms,''
  in \emph{{STOC}}, 2002, pp. 380--388.

\bibitem{LSH98}
P.~Indyk and R.~Motwani, ``Approximate nearest neighbors: Towards removing the
  curse of dimensionality,'' in \emph{{STOC}}, 1998, pp. 604--613.

\bibitem{zhong2016tracking}
Y.~Zhong, S.~Liu, X.~Wang, J.~Xiao, and Y.~Song, ``Tracking idea flows between
  social groups.'' in \emph{AAAI}, 2016, pp. 1436--1443.

\end{thebibliography}

\end{document}